# The principle and applications of Fourier back plane imaging*

*Yurui Fang*

(School of Physics, Dalian University of Technology, Dalian, 116024, Email: yrfang@dlut.edu.cn)

**Abstract**

Fourier back plane (FBP) imaging technique has been widely used in the frontier research of nanophotonics. In this paper, based on the diffraction theory and wave front transformation principle, the FBP imaging basic principle, the setup realization and the applications in frontier research are introduced. The paper beginnings with the primary knowledge of Fourier optics, combining with the modern microscope structure to help to understand the Fourier transformation principle in the advances of nanophotonics. It can be a reference for experimental teaching and researching.

## Contents

## 1 Introduction

Fourier transform optics is an old direction but with important modern applications in optics which is an ancient but novel subject. As the developing of micro/nano- fabrication techniques and nano-optics, and the research requirement on small scaled objects like quantum dots et al, Fourier transform microscopy is re-boosting based on the modern technique since Lie's work.[1] The purpose of the paper is introducing the Fourier transform microscopy and its applications in frontier research based on the basic Fourier transform optics principle.

The principle of Fourier transform optics is introduced first, and then the realization in experiments based on the infinity optical system is following. After that those fundamental principles are used to analyze the applications in high order scattering imaging with high resolution, wave vector measurement in surface waves, emission/scattering direction measurement in small particle or molecule systems and the k- space imaging for photonic crystals. Last but not least, the low cost and simple Fourier transform imaging setup that can be realized in an undergraduate student lab is discussed. In focusing the basic principle learned in university and the applications with different techniques, we only focused on the basic phenomenon on Fourier transform but not the deep mechanism of the studied system so as not to lose the center point.

As the studied objects are always much smaller than the front lens size of the objective, to avoid complicated mathematic calculations, we assume the objective transmission window function is 1. Only the knowledge of Fresnel diffraction, Fraunhofer diffraction and the phase transformation by the lens is used. The angular spectrum theory is avoided in the following. And only monochromatic light is considered in the whole paper; the non-monochromatic light is approximately considered as the linear sum of monochromatic light with different wavelengths.

## 2 The principle of Fourier transform[2]

According to the basic principle，the integral form of 2 dimension (2D) Fourier transform of a function $U$ is

$$F_U(f_x, f_y) = \iint\limits_{-\infty}^{+\infty} U(x,y) e^{-j2\pi(f_x x + f_y y)} dx dy \quad (1).$$

And the inverse transformation is

$$U(x,y) = \iint\limits_{-\infty}^{+\infty} F_U(f_x, f_y) e^{j2\pi(f_x x + f_y y)} df_x df_y \quad (2)$$

The convolution relation of Fourier transform is: if

$$U_1(x,y) = \iint\limits_{-\infty}^{+\infty} U_2(\xi, \eta) h(x-\xi, y-\eta) d\xi d\eta \equiv U_2(\xi, \eta) \otimes h(x,y) \quad (3)$$

Then the Fourier transform satisfies

$$F_{U1}(f_x, f_y) = H(f_x, f_y)F_{U2}(f_x, f_y) \quad (4)$$

where $H(f_x, f_y)$ is the Fourier transform of $h(x, y)$:

$$H(f_x, f_y) = \iint\limits_{-\infty}^{+\infty} h(x,y)e^{-j2\pi(f_x x + f_y y)}dxdy \quad (5)$$

## 2.1 A diffraction screen is the superposition of cosine gratings

As the studied object or the diffraction screen is composed with small molecules or atoms whose sizes are much smaller than the light wavelength, they can be considered continuum. Then we assume that the spatial character of any (1 dimensional, 1D) diffraction screen can be decomposed into the superposition of cosine gratings with different frequencies:

$$t(\mathrm{x}) = t_0 + \sum_{n \neq 0} t_n e^{j2\pi f_n x} \quad (6)$$

The essence of diffraction integration is the linear superposition of pulse responses. Therefore, the whole screen can be studied with Fourier transformation. 2D screens are just a direct expending of 1D condition.

## 2.2 Fresnel diffraction and Fraunhofer diffraction as a principle of frequency division

Fresnel approximation and Fraunhofer approximation is equivalent as the paraxial approximation.[2] From Huygens-Fresnel principle and Kirchhoff integral formula, we know that the field distribution

（$r_0 \gg \lambda$）of a plane diffractive screen $U(\xi, \eta)$ under Fresnel approximation （$z^3 \gg \frac{\pi}{4\lambda}[(x - \xi)^2 + (y - \eta)^2]_{max}^2$）is

$$U(x,y) = \frac{e^{jkz}}{j\lambda z}e^{j\frac{k}{2z}(x^2+y^2)}\iint\limits_{-\infty}^{+\infty}[U(\xi,\eta)e^{j\frac{k}{2z}(\xi^2+\eta^2)}]e^{-j\frac{2\pi}{\lambda z}(x\xi+y\eta)}d\xi d\eta \quad (7).$$

So the Fresnel result is the product of a phase factor (related to $x^2 + y^2$) and the Fourier transform of the complex field close to the screen (related to phase factor $\xi^2 + \eta^2$). The Fresnel diffraction can be treated as a Fourier transform of the field close to the diffractive object.

The diffraction formula under Fraunhofer approximation （$z \gg \frac{k}{2}(\xi^2 + \eta^2)_{max}$）is

$$U(x,y) = \frac{e^{jkz}}{j\lambda z}e^{j\frac{k}{2z}(x^2+y^2)}\iint\limits_{-\infty}^{+\infty}U(\xi,\eta)e^{-j\frac{2\pi}{\lambda z}(x\xi+y\eta)}d\xi d\eta \quad (8).$$

Compared with Fresnel diffraction, the $e^{j\frac{k}{2z}(\xi^2+\eta^2)}$ factor disappears. One can see that excepting a $x^2 + y^2$ phase factor, the Fraunhofer diffraction is the Fourier transform of the diffractive object. The transformed frequencies are

$$f_x = x/\lambda z \quad (9a)$$

$$f_y = y/\lambda z \quad (9b).$$

Because it is very rigor to satisfy the Fraunhofer diffraction conditions, in the lab one usually need lenses to realize the experiment. So the Fraunhofer diffraction with lenses is a Fourier transform. The Fraunhofer diffraction thus can be used to realize (special) frequency division.

### 2.3 The transmission function of thin lens

Consider the phase delay of a common lens shown in Figure 1:

$$\phi(x, y) = kn\Delta(x, y) + k[\Delta_0 - \Delta(x, y)] \quad (10)$$

$$t_l(x, y) = exp[jk\Delta_0]exp[jk(n-1)\Delta(x, y)] \quad (11)$$

$$\Delta(x, y) = \Delta_0 - R_1 \left( 1 - \sqrt{1 - \frac{x^2 + y^2}{R_1^2}} \right) + R_2 \left( 1 - \sqrt{1 - \frac{x^2 + y^2}{R_2^2}} \right) \quad (12)$$

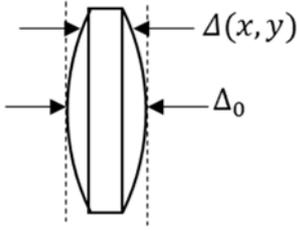

**Figure 1. Schematic illustration for calculating the optical length function of a thin lens. The radiuses of the left and right sphere surfaces are $R_1$ and $R_2$, respectively. When the light comes from left and goes out from right, we use $R_1 > 0$ and $R_2 < 0$.**

Because only the intensity is considered in the end, so the constant phase factor $\Delta_0$ is ignored. Under the paraxial condition,

$$\sqrt{1 - \frac{x^2 + y^2}{R_i^2}} \approx 1 - \frac{x^2 + y^2}{2R_i^2} \quad (13)$$

$$\Delta(x, y) = \Delta_0 - \frac{x^2 + y^2}{2} \left( \frac{1}{R_1} - \frac{1}{R_2} \right) \quad (14)$$

$$\frac{1}{f} = (n-1) \left( \frac{1}{R_1} - \frac{1}{R_2} \right) \quad (15)$$

$$t_l(x, y) = exp[jk(n-1)\Delta(x, y)] = exp \left[ -jk\frac{x^2 + y^2}{2f} \right] \quad (16)$$

If the distribution of the incident plane wave is 1, the transmission field is:

$$U_l(x, y) = exp \left[ -j\frac{k}{2f}(x^2 + y^2) \right] \quad (17)$$

which is a spherical wave.

### 2.4 The Fourier transform property of thin lens

Considering an optical path shown as Figure 2a, an incident objective screen (with a transmission function $t_o(x, y)$) is put closely before a thin lens parallel. The screen is illuminated with a plane wave (amplitude is A) perpendicular to the screen, then the field after the thin lens is

$$U_l(x, y) = At_o t_l = At_o \exp\left[-jk\frac{x^2 + y^2}{2f}\right] \quad (18)$$

Using Fresnel diffraction formula, ignoring the constant phase factor $e^{jkf}$, the field on the back-focus plane ($z = f$) of the lens is

$$U_f(u, v) = \frac{e^{j\frac{k}{2f}(u^2+v^2)}}{j\lambda f} \iint\limits_{-\infty}^{+\infty} U_l(x, y)e^{j\frac{k}{2f}(x^2+y^2)}e^{-j\frac{2\pi}{\lambda f}(xu+yv)}dxdy$$

$$= \frac{e^{j\frac{k}{2f}(u^2+v^2)}}{j\lambda f} \iint\limits_{-\infty}^{+\infty} At_o(x, y)e^{-j\frac{2\pi}{\lambda f}(xu+yv)}dxdy \quad (19).$$

From the expression one can see that the field distribution is exactly the same as the Fraunhofer diffraction formula 8. This is also the reason that one use lenses to realize Fraunhofer diffraction in the lab. If the lenses are used, the image on the focus plane can be considered as the image of the object from infinity distance, which satisfies the Fraunhofer diffraction condition. One should notice that under this condition, because of the quadratic phase factor, the Fourier transform relation is uncertain yet.

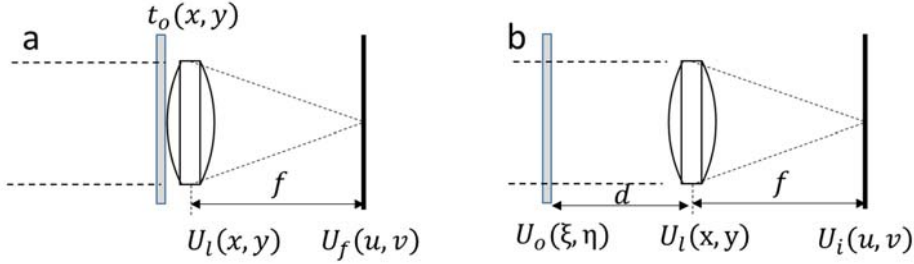

Figure 2. Optical path illustration for the back-focus plane imaging.

Now considering the optical path shown in Figure 2b, an incident objective screen (with a transmission function $t_o(\xi, \eta)$) is put before a thin lens parallel with distance $d$. The screen is illuminated with a plane wave (amplitude is A) perpendicular to the screen. For convenience, assume $F_O(f_\xi, f_\eta)$ is the Fourier transform frequency spectrum of the diffraction screen field $U_o = At_o(x, y)$; $F_l(f_x, f_y)$ is the Fourier transform frequency spectrum of the field $U_l(x, y)$ just before the lens that comes from the screen. Then ignoring the constant phase factor $e^{jkf}$, the field on the back-focus plane ($z = f$) of the lens is

$$U_i(u, v) = \frac{e^{j\frac{k}{2f}(u^2+v^2)}}{j\lambda f} \iint\limits_{-\infty}^{+\infty} U_l(x, y)e^{-j\frac{2\pi}{\lambda f}(xu+yv)}dxdy = \frac{e^{j\frac{k}{2f}(u^2+v^2)}}{j\lambda f}F_l(f_x, f_y) \quad (20),$$

where $f_x = \frac{u}{\lambda f}, f_y = \frac{v}{\lambda f}$.

From Fresnel diffraction formula, we have

$$U_l(x,y) = \frac{e^{jkz}}{j\lambda z} e^{j\frac{k}{2z}(x^2+y^2)} \iint\limits_{-\infty}^{+\infty} [U_o(\xi,\eta) e^{j\frac{k}{2z}(\xi^2+\eta^2)}] e^{-j\frac{2\pi}{\lambda z}(x\xi+y\eta)} d\xi d\eta$$

$$= \iint\limits_{-\infty}^{+\infty} U_o(\xi,\eta) [\frac{e^{jkd}}{j\lambda d} e^{j\frac{k}{2d}((x-\xi)^2+(y-\eta)^2)}] d\xi d\eta = U_o(\xi,\eta) \otimes h(x,y) \quad (21),$$

where $h(x,y) = \frac{e^{jkd}}{j\lambda d} e^{j\frac{k}{2d}(x^2+y^2)}$, and its Fourier transform is

$$H(f_x, f_y) = \iint\limits_{-\infty}^{+\infty} h(x,y) e^{-j2\pi(f_x x + f_y y)} dx dy = e^{jkd} e^{-j\pi\lambda d(f_x^2 + f_y^2)} \quad (22)$$

where $f_x = \frac{u}{\lambda d}, f_y = \frac{v}{\lambda d}$. So

$$F_l(f_x, f_y) = H(f_x, f_y) F_O(f_\xi, f_\eta) \quad (23).$$

Substitute it into formula 20, we have

$$U_i(u,v) = \frac{e^{j\frac{k}{2f}(u^2+v^2)}}{j\lambda f} H(f_x, f_y) F_O(f_\xi, f_\eta) = \frac{e^{j\frac{k}{2f}(u^2+v^2)}}{j\lambda f} e^{jkd} e^{-j\pi\lambda d(f_x^2+f_y^2)} F_O(f_\xi, f_\eta)$$

$$= g(u,v) F_O(f_\xi, f_\eta) \quad (24),$$

where

$$g(u,v) = \frac{e^{j\frac{k}{2f}(1-\frac{d}{f})(u^2+v^2)}}{j\lambda f} \quad (25).$$

The constant phase factor $e^{jk(d+f)}$ is ignored during the calculation.

From formula 24 one can see that the field on the back-focus plane is the product of the Fourier transform of the object screen $U_o(\xi,\eta)$ and a quadratic phase factor $g(u,v)$. When $d = 0$, formula 24 comes back to the Fraunhofer diffraction formula 8 and the screen lens transmission formula 19. When $d = f$, $g(u,v)$ becomes a complex term, which doesn't affect the field distribution on the back-focus plane. Hence, at this condition, the field on the back-focus plane is totally the Fourier transform of the object screen field $U_o(\xi,\eta)$. When $d = \infty$, the constant phase factor $e^{jk(d+f)}$ should be considered. If $u^2 + v^2 \ll f^2$, $g(u,v) \approx \frac{e^{jkd}}{j\lambda f}$ is a constant phase factor as well. So the relation now also satisfies the Fourier transform relation, which will be discussed later.

## 2.5 An example: frequency spectrum analysis and cosine gratings

From section 2.1 we know that cosine gratings can be used to decompose different diffractive screens as basic diffraction elements. 1D cosine grating is used as a diffraction screen in the following part as an example to show the process. Assume the transmission function of the cosine grating screen is

$$t_{cos}(x, y) = t_0 + t_1\,cos(2\pi f_x x + \psi_0) \quad (26).$$

When it is illuminated with a monochromatic plane wave (wave front is $U_0$) perpendicular. Set $\psi_0 = 0$, after the screen, the wave front is

$$U_{cos}(x, y) = U_0 t_{cos}(x, y) = U_0(t_0 + t_1\,cos(2\pi f_x x))$$

$$= U_0 t_0 + \frac{1}{2}U_0 t_1 e^{j(2\pi f_x x)} + \frac{1}{2}U_0 t_1 e^{-j(2\pi f_x x)}$$

$$= U_0 t_0 + \frac{1}{2}U_0 t_1 e^{jk(f_x \lambda x)} + \frac{1}{2}U_0 t_1 e^{-jk(f_x \lambda x)}$$

$$= U_0 t_0 + \frac{1}{2}U_0 t_1 e^{jk_x x} + \frac{1}{2}U_0 t_1 e^{-jk_x x} = \widetilde{U}_0 + \widetilde{U}_1 + \widetilde{U}_{-1} \quad (27).$$

If the screen is on the front focus plane of the lens, the field on the back-focus plane of the lens is (nothing to do with y direction)

$$U_i(u, v) = \frac{1}{j\lambda f}F_O(f_\xi, f_\eta) = \iint\limits_{-\infty}^{+\infty}(\widetilde{U}_0 + \widetilde{U}_1 + \widetilde{U}_{-1})e^{-j2\pi(f_x x + f_y x)}dxdy$$

$$= 2\pi\delta(u - 0)\delta(v - 0) + \pi\delta(u + k_x)\delta(v - 0) + \pi\delta(u - k_x)\delta(v - 0) \quad (28).$$

From the expression one can see that the patterns on the back-focus plane are three bright points. One is on the lens axis and the other two are on two sides of the axis (Figure 3). Where $k_x = 2\pi f_x = 2\pi/d_x$, which reflects the period of the screen and the wave vector on x direction which $k_y = 0$.

Other periodic functions can be treated as the superposition of many cosine gratings with different periods, each of which has corresponding frequency point on the back focus plane so as to divide the frequencies.

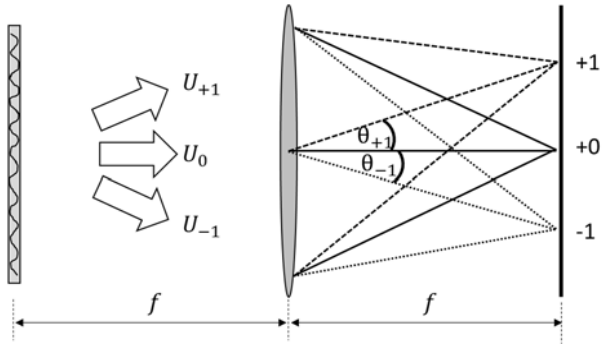

**Figure 3. Schematic illustration of 1st order diffraction of a cosine grating under normal illumination of a monochromatic plane wave.**

## 2.6 Analyzing the diffraction limit and near field optics from the point view of diffraction and thin lens Fourier transform[3]

From expression 27 one can see that $\widetilde{U}_0$, $\widetilde{U}_1$, $\widetilde{U}_{-1}$ represent plane waves respectively (Figure 4a). $\widetilde{U}_0$ is an outgoing plane wave in normal direction which is the zero order wave without component in $x$ direction. $\tilde{U}_1$ is the plane wave propagating oblique upward and $\tilde{U}_{-1}$ propagating oblique downward. The oblique angle is

$$\sin\theta_{\pm1} = \pm f_x\lambda = \pm\frac{\lambda}{d_x} = \frac{\pm n\lambda}{nd_x}, \quad n = 1,2,3,\dots \quad (29)$$

Those three plane waves will be focused on three points after the lens, which are the diffractive patter of cosine gratings. Formula 29 is exactly the 1$^{st}$ order diffraction formula of Fraunhofer diffraction, which can also be treated as higher order diffraction of gratings with period $nd_x$.

Let $f_0 = 1/\lambda$ represents the spatial frequency, the spatial structures on the diffraction screen can be classified into three orders: $f_x \ll f_0$ (low frequency structures), $f_x \leq f_0$ (high frequency structures), $f_x > f_0$ (hyperfine structures). Considering $f_x > f_0$, $\sin\theta_{\pm1} = \pm f_x\lambda = \pm\frac{f_x}{f_0} > 1$, where the solutions of $\theta_{\pm1}$ are complex number. Therefore, the diffraction field becomes evanescent field (Figure 4b). For clarity, the expression of the plane wave propagating in the direction is

$$U_{+1} = U_0 e^{j(k_x x + k_y y + k_z z)} \quad (30),$$

where

$$k_x^2 + k_y^2 + k_z^2 = k^2 = (2\pi/\lambda)^2 = (2\pi f_0)^2 \quad (31).$$

Meanwhile, on the gratings output plane one has

$$k_x = 2\pi f_x, \quad k_y = 0 \quad (32).$$

So

$$k_z = \sqrt{k^2 - (k_x^2 + k_y^2)} = 2\pi\sqrt{f_0^2 - f_x^2} = 2\pi f_0\sqrt{1 - \left(\frac{f_x}{f_0}\right)^2} = jk_z' \quad (33),$$

where $k_z' = 2\pi f_0\sqrt{\left(\frac{f_x}{f_0}\right)^2 - 1}$ is a real number. This kind of field will decay very fast in z direction, which is called evanescent wave. This also means that the optical information from hyperfine structures cannot propagate to far field through the lens. Hence, from this point of view, the resolving limit of traditional optical microscopes must be in the order of wavelength.

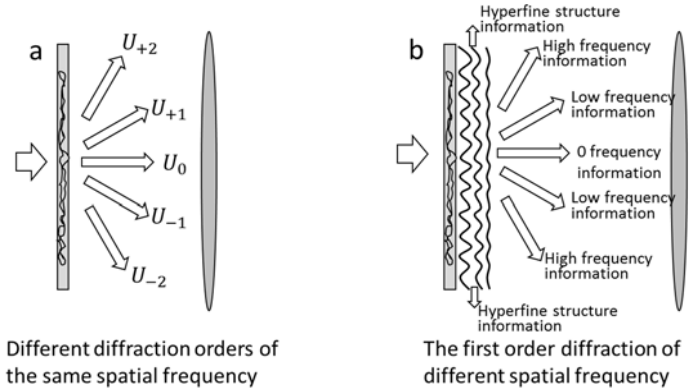

Figure 4. Schematic illustration for the frequency division by diffractive screen under monochromatic plane wave illumination.

During the imaging process, usually the caliber of the objective has limited size, so the higher order diffraction waves cannot be collected so as that some information will be lost naturally. Maybe some says that one can break through the diffraction limit by increasing the caliber of the objective. However, from the discussion above we know that with the increasing of the caliber, more and more higher order diffraction waves and the waves from the structure approaching the wavelength size can be collected, but one cannot break though the diffraction limit because the evanescent waves cannot be collected. If higher refractive index medium is added between the objective and the sample (like the oil immersed objective), the diffraction angle $\theta_{\pm 1}$ will decrease. So increasing the numerical aperture $NA = nsin\theta_{\pm 1}$ (equivalent to increase $f_x$) will make more hyperfine structure diffraction light be collected, which will increase the resolution. However, the refractive index of the matching oil cannot change too much. So from the point of Fourier transform, the diffraction limit cannot be breakthrough in this way. The main methods to increase the resolution of non-scanning optical microscope (the microscope based on point spread function and data processing or scanning microscope in not included here in this paper) are decreasing the wavelength $\lambda$ (increasing $f_0$), in which using high order harmonic wave to realize super resolution is one of common ways. Because the diffraction field of hyperfine structures are evanescent waves which cannot propagates to far field, the only way to detect the optical information of such structures is the scanning near field techniques (the microscope based on point spread function and data processing can only give the far field information).

### 3 Setups for Fourier back plane (FBP) imaging

Besides the traditional Fourier transform with lenses, The FBP technique is re-boosting in the field of plasmonics and nanophotonics, which is a natural consequence of micro-/nano- fabrication techniques. To observe such samples, a microscope should be utilized. Then in the following we first take a look at the FBP setups based on modern microscope systems.

### 3.1 The infinity optical systems satisfies the Fraunhofer diffraction approximation and thin lens conditions

**Infinity optical system**

The modern microscope used in research is different from the old microscopes learned in school after several generations' improvement. In the old microscope (Figure 5a) samples are put between the focus length (f) and 2f. There will be a magnified inverted real image inside the barrel. Then a CCD or eyepiece is used to detect or observe the image. The magnification of the microscope is the magnification of the eyepiece (or CCD) multiplying that of objective. In such kind of design, the length of the barrel is fixed and the standards are different for different brands or manufacture. Usually it is 160 mm, 180 mm or 200 mm. So the disadvantage is quite obvious: if one wants to add some other components like polarizers, the length of the barrel must be re-designed, which is very troublesome. The modern microscopes are based on infinity optical system (Figure 5b). In infinity optical system, the samples are on the front focus plane of the objective. The light from each point of the sample will be a parallel beam after the objective. Then a tube lens is used to focus the parallel beams of different angles on the back-focus plane of the tube lens forming a magnified inverted real image. Then a CCD or eyepiece is used to detect or observe the image. In the middle of the barrel between the objective and the tube lens, because the beam is parallel, any component with parallel sides can be added into the optical path conveniently without affecting the barrel or the structure. As the output beam from the objective is parallel, the barrel can be infinitely long principally, which gives the name of the system as infinity optical system. The focus lengths are fixed (160 mm, 180 mm or 200 mm), which matches the old design of the old system for each manufacture. The magnification before eyepiece is the focus length of the tube lens divided by the focus length of the objective.

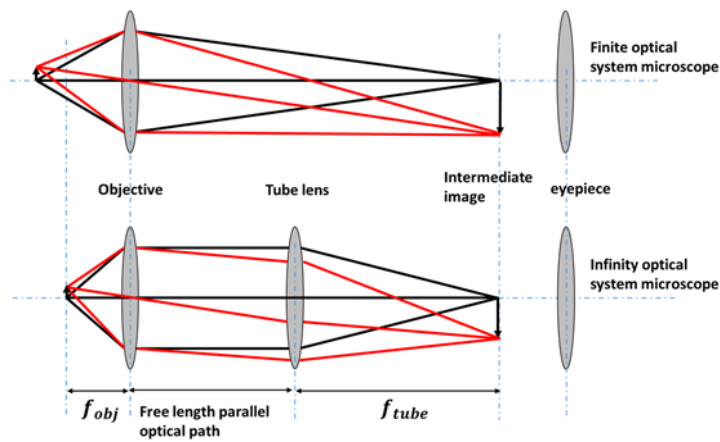

**Figure 5. The schemes of finite and infinity optical microscopes.**

**The objective and tube lens in infinity optical system**

The objectives used in infinity optical systems (Figure 6) have different designs like aplanatic aberration, achromatic aberration, coma aberration or the combination of those. The ingenious design fits the paraxial condition. Right now we haven't found any reference to directly do the optical Fourier transform analysis for the formulation of such complex system composed with complex lens groups as in section 2. It is maybe in the secret files in the manufactures. From the light path of the patent figure (Figure 6b) of such objective, one can treat the objective as a black box. If only the entry and exit pupils are considered, the objective satisfies the paraxial condition of thin lens. Even though there is no exact calculations given in the patent, Kurvits et al studied the objectives of different brands and different specifications by geometric optics method with the parameters given in the patents and disassembled objectives (Figure 7).[4] They showed that when the $NA$ is larger than 1.3, the distortion would be quite obvious. And higher $NA/n_{immerge}$ will cause higher distortion. Generally, in Fourier transform imaging, objectives with lower magnification and higher $NA$ should be used. Smaller magnification will make larger FBP image and the image will not blur with increased focus length. In practice, one needs certain magnification, so the objectives with higher matching medium like oil objectives are used which also have larger $NA$ and exit pupil. When building the FBP setups, lenses with long focus length are also a better choice. With the decreasing of the light emitting object, the distortion becomes smaller.

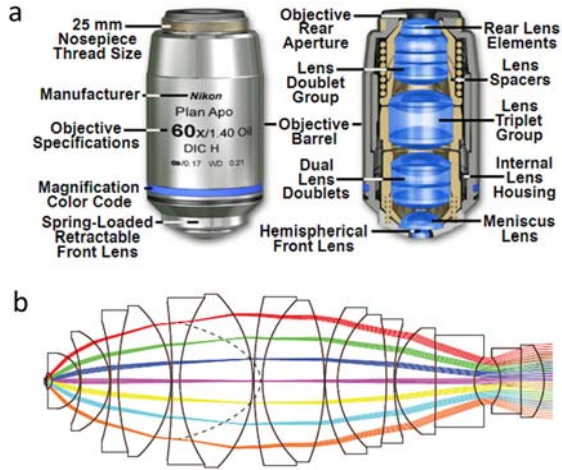

**Figure 6. (a) A kind of infinity aplanatic achromatic plane field oil immersed objective. (Figures are reprinted with permission from ref. 5, Nikon)[5]. (b) The inside optical path of an infinity objective (Figure is reprinted with permission from ref. 6)[6].**

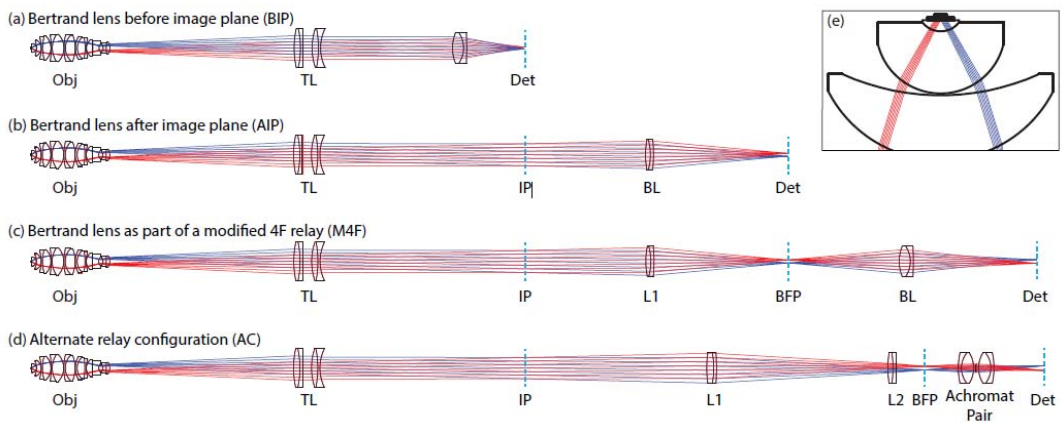

**Figure 7. The study of FBP distortion with geometry optics method (Figures are reprinted with permission from ref. 4, OSA)[4].**

In scientific researches, to directly measure the emission angle of the samples, oil immersed aplanatic achromatic plane field objectives with high magnifications (>50x) are usually used. Though Kurvits' research[4] shows that the Fourier transform result by low magnification plane field objective is better, in practice, only the small area from the center of the field of view is concerned. So in such measurement, the optical path satisfies the approximation conditions we discussed above.

**Illumination and imaging**

The bright field illumination is through Cole lighting system which is approximately parallel incident light. The emission for the dark field illumination can be treated as direct emission from points satisfying the Huygens-Fresnel secondary condition which can be calculated directly with the Fresnel approximation and formula 7 (from another point of view, the dark field illumination can be treated as spherical wave illumination, if the object screen is on the front focus plane of the lens, the field on the back focus plane is also a strict Fourier transform of the object screen).

In the practical measurement with the infinity microscope, the sample is on the front focus plane of the objective, so the transform is a Fourier transform in the back focus plane of the objective under the illumination. However, the conjugate focus point of the objective is inside of the objective barrel, which is almost impossible used for FBP imaging here, usually extra relay optical path is needed. The common method is to use thin lens to do the FBP for the inverted real image on the back focus plane (usually outside of the microscope out port) of the tube lens.

### 3.2 The setups for FBP

In practical scientific researches, according to the position of the Bertrand lens, there are four kinds of configurations for the FBP setup (Figure 7)[4]. But for convenience and the magnification of the FBP image, 4f systems are used (even though Kurvits' research shows that the BIP and AIP configurations are better) in which by changing the focus length of the last lens one can collect the FBP image or real space image. A typical scheme for such 4f system is shown in Figure 8. Lens 1 and Lens 2 have the same focus length (usually for broadband wavelengths). The real image by the tube lens is at position 3 (the back focus plane of tube lens), which is also at the front focus plane of Lens 1. The image at position 3 is formed by the focusing of tube length, so it can be treated as a spherical wave illuminated diffraction screen. The image on the back focus plane of lens 1 is a strict Fourier transformation of the image 3. Now if the front focus plane is overlapping with the back focus plane of lens 1, then on the back focus plane of lens 2 the image will be the Fourier transformation of Fourier plane I, which is a real image. We can detect this image with a CCD at here. If we replace lens 2 into a lens 2' with half focus length with a wheel, then both the Fourier plane I and image plane II are at the 2f plane of lens 2'. At the CCD position, it will form an inverted image of plane I, which is still the FBP image of image plane 3.

During the imaging process, we can see that for monochromatic light emission, the transform is strict Fourier transform. But in the real experiments with white light imaging, the Bertrand lens cannot be corrected in the whole wavelength band and the Fourier transform will have some deviation. So in the real experiments, the lenses are chosen with the focus lengths are calibrated at the central wavelength of the experiments to reduce the deviation. And lenses with long focus lengths can also reduce the dispersion distortion.

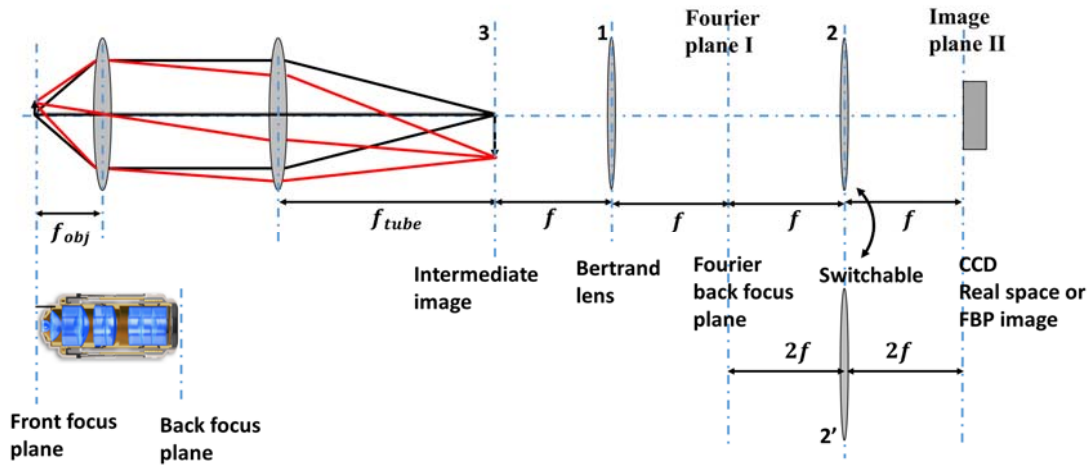

**Figure 8. A typical scheme of FBP setup. The picture on the lower left is from Nikon's web page(Figure is reprinted with permission from ref. 5，Nikon)[5]。**

Another issue is that the Fourier transform in above setup is a strict Fourier transformation of the image on plane 3. But the object studied is the sample at the front focus plane of the objective. Then from the sample plane to the image plane of the tube lens, if the setup guarantees that the image from the tube lens plane 3 reflects the optical properties of the sample strictly? From the discussion in last section we know that the paraxial condition and the thin lens condition are both satisfied for the objective. In such infinity optical system, the tube lens is usually composed with two glued lenses. The angle of the incoming light for the tube lens (also the same as the outgoing light angle of the objective rear pupil) is quite small, which satisfies the paraxial condition. When the tube lens imaging, because the exit light of the objective is parallel beams (for each single point on the sample, it satisfies the point spread function and the linear response condition). Hence, inside the barrel of such infinity optical system, the optical path is a Fraunhofer diffraction of the secondary point sources on the back focus plane of the objective. What we do is that we just image the Fraunhofer diffraction pattern with a tube lens. So the image of the tube lens is a strict Fourier transform of the objective back focus plane image (Fourier transform of the object plane). From another point of view, this condition can be treated as $d = \infty$ in formula 24, which also satisfies the Fourier transform relation. During the process, to get better image quality (they usually never study the Fourier transform properties of the microscope), the manufactures design larger rear pupil and restrict exit angle at a quite small value so as to make more information transmit to the entry plane of the tube lens.

## 4 The applications of FBP in frontier research

The theory of Fourier transform optics has already been quite complete and applied broadly since 60' in last century. Just like the re-boosting of plasmonics and nanophtonics, the Fourier transform optics also benefits from the developing the nanotechnology and re-developed since 1990s'. The rising of nanophotonics brings new phenomenon in the diffractive limit scale. In the following we the applications of FBP in frontier research are introduced based on the above analysis.

### 4.1 Spatial filtering and *k*-space reconstruction and imaging

From the textbook of university course we know that one application of traditional Fourier transform optics is spatial filtering and image differential, which is a direct result of Abel's principle. One of the relative applications in advanced sciences is also image differential, which makes the image clearer.

In the optics course we know that when object is on the front focus plane of objective, there are lot of diffraction spots on the FBP. If some of the high order spots are blocked, the image will be blur. While if the low order spots are blocked, the image edges will be very sharp, just like doing differential to the picture in the software. The principle can be used for programmable image processing in the frontier research. As shown in Figure 9[7], the object is on the front plane of the objective, L1 is the tube lens, and the image by the tube lens is in the image plane 1. The light transmits through lens L2, there will be a Fourier image on the back focus plane of L2, which can be collected at the position LCOS. In this work there is a programmable spatial modulator before LCOS, whose pixels can be controlled independently with computer to reflect the light or not. Because the modulator is on the FBP of L2, different order of diffraction spots can be controlled to be reflected, like in Figure 9b, the lower order diffraction in the middle of the plane is absorbed, and higher order diffraction in outer place is reflected. The reflected light is polarized and reflected again by the NPBS, and then re-imaged (Fourier transform again) by lens 3 and detected by a CCD. The whole process is like a high pass spatial filter which is used for image differential. Therefore, the image only has bright edge.

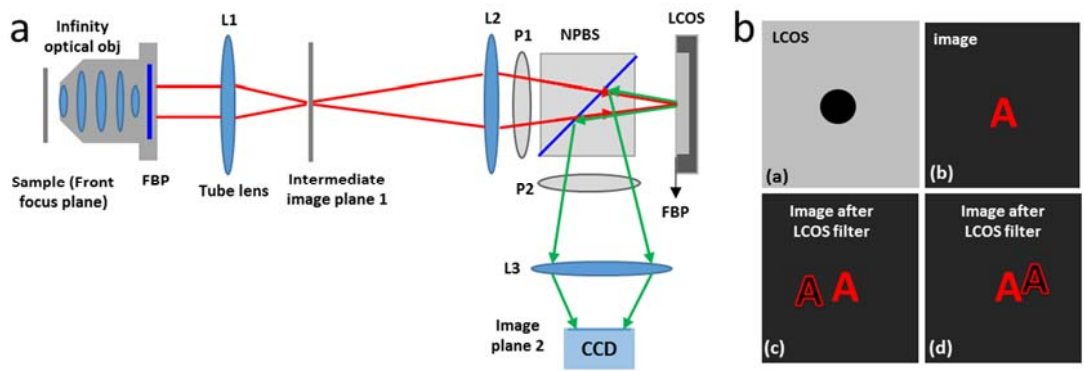

**Figure 9. (a) Schematic illustration of a programmable Fourier spatial filter microscope described in Ref. 7. (b) Schematic illustration of imaging by the microscope in Ref. 7[7].**

Another technique developed in recent years called Fourier ptychographic microscopy (FPM) is also use Fourier transform spatial filter to realize very high resolution image with low magnification objective. In the beginning this technique is more depending on the computer algorithms to reconstruct the image with the information detected by CCD on image plane[8]. Then recently Tian et al use the property that the Fourier back plane of the objective is inside the rear pupil, which is naturally a spatial filter for the FBP. A special algorithms is used to calculate the exit pupil transmission function to reconstruct the image[9]. As Figure 10 shown, the LED arrays will illuminate the sample from different angles. According to the displacement-phase shift principle, we know that it is equivalent to add incident phase differences on the sample. So the Fraunhofer diffraction spots will shift in $k$-space (Fourier space). But the rear pupil has fixed position, so it will filter different diffraction orders for different illumination angles by LED. After the images are acquired, they can calculate the amplitude and phase information with Fourier transform with the LED position information. The information will be input information for next round of calculation with switched LED lightening up. The algorithms is self-consistent, which will output the final reconstructed high resolution image. We will not focus on the reconstruction algorithms but on the Fourier transform essence for high resolution in this technique. In the words of the paper, the reason to get high resolution is equivalent to have high *NA*. Now recall the knowledge in section 2, we know that when the sample is illuminated with tilted light, the 0-order diffraction spot will shift, then the high order diffraction light from one side (that cannot enter the objective under normal illumination) will change angle and goes into the objective with smaller angle shown as in Figure 10a. Meanwhile, part of the diffraction of hyperfine structures (which is evanescent wave under normal illumination) will start to emit light into far field, which has the opportunity to enter the objective. So this is the reason that one can have high resolution image with this technique from Fourier transform optics analysis. Similar works can be referred to ref. [8, 10-11].

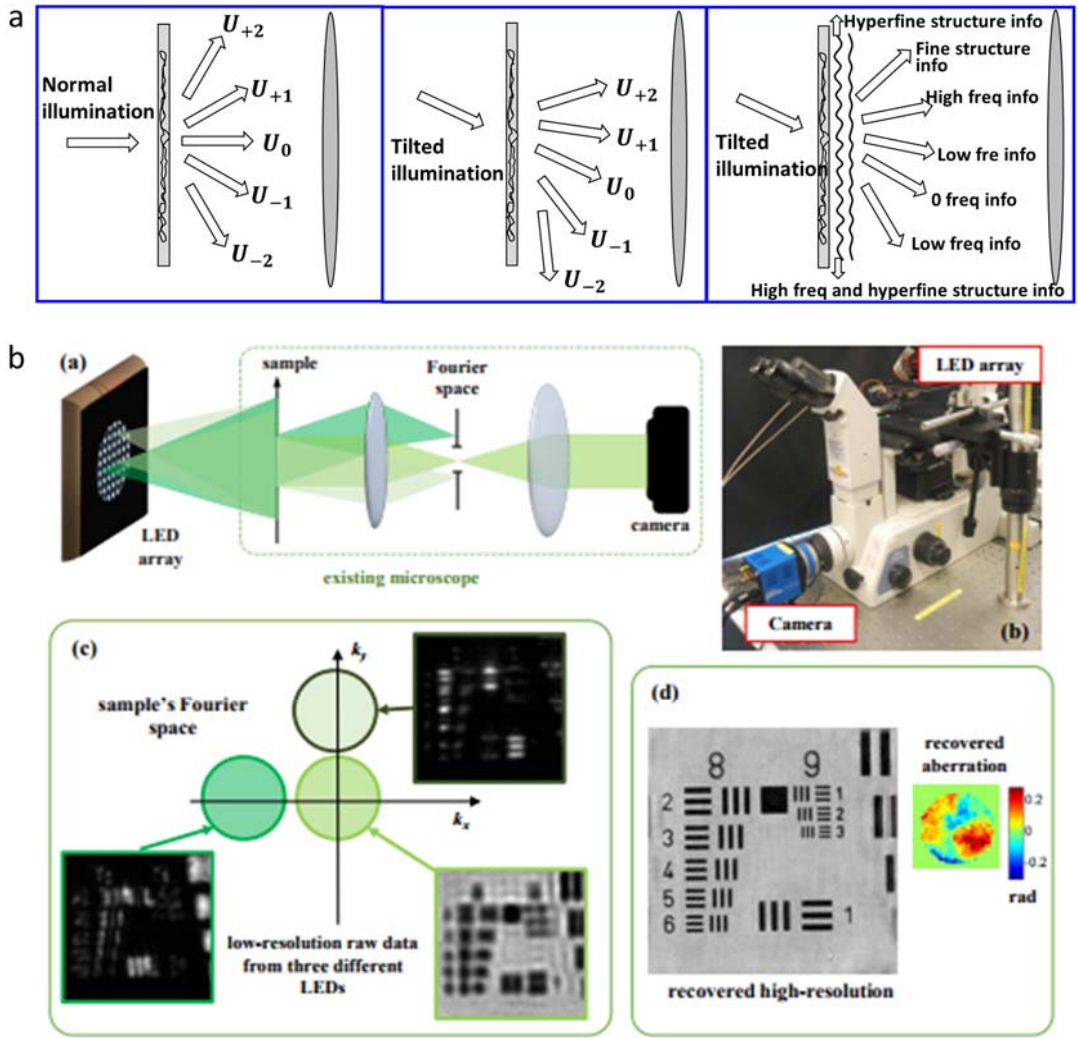

**Figure 10. (a) Scheme of different diffraction orders under normal or tilted illumination. (b) A high resolution construction microscope with multi-point illumination: a. principle scheme; b. experiment setup; c. corresponding k-space positions under different LED illumination; d. the final high resolution image. (Figures are reprinted with permission from ref. 9，OSA)[9].**

### 4.2 Leakage Radiation Microscopy and its applications: waveguide modes measurement

Form section 2 we know that the Fourier transformed spatial frequency $f_x$ of the object screen actually reflects the wave vector ($k_x = 2\pi f_x$) of the diffractive light in the direction perpendicular to the normal direction. Taking the cosine grating as an example: the wave satisfies the periodic boundary condition in the screen plane which makes only the light whose wave vector matches the periodic condition exist. On the other hand, once there are periodic waves on the front focus plane and can propagate to far field, one can use FBP to image the frequency. This kind of FBP setup (section 3) is usually called Leakage radiation Microscope (LRM), which comes from the optical reciprocity principle. It has been known that when a light beam shining from optically dense medium to thin medium, if the incident angle is larger than the critical angle, total reflection will happen and evanescent wave will be excited on the interface. The wave vector of the evanescent wave is the same as the parallel component of the incident wave. Conversely, when there is a propagating evanescent wave on the interface, there will be a corresponding emission beam goes out in the optically dense medium, whose wave vector's parallel component is the same as the evanescent wave vector. The outcome beam is called leakage mode. The principle is shown in Figure 11a. From the discussion in section 2 we know that the evanescent wave actually reflects the spatial characters of the interface. FBP image works with this phenomenon. So such LRM is usually built with an inverted microscope equipped with oil immersed objective (of course there are LRM built with upright microscope), which directly matches the refractive index of the substrate.

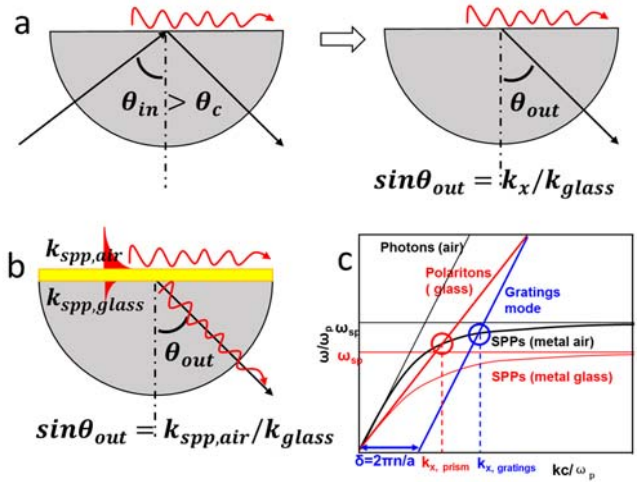

**Figure 11. (a) Optical reciprocity principle. (b) Leakage mode of SPP on Au film. (c) SPP dispersion relation of metal film / dielectric interface[12].**

The FBP imaging for wave vector is commonly used for surface plasmon polartons (SPPs) measurement. The principle is shown in Figure 11b. The wave vector of SPPs on the thin metal film evaporated on glass substrate $k_{spp,medium} = \frac{\omega}{c} \sqrt{\frac{\varepsilon_{metal}\varepsilon_{medium}}{\varepsilon_{metal}+\varepsilon_{medium}}}$. There are two SPP waves on the two interfaces ($k_{spp,air}$ and $k_{spp,glass}$). The wave vector on air/film interface ($k_{spp,air}$) is a little large than the light wave vector in air (usually smaller than $1.1 k_0 = 1.1 * \frac{\omega}{c}$) and $k_{spp,air} > k_{glass} = 1.5 * \frac{\omega}{c}$. According to the principle of leakage mode, $k_{spp,glass}$ cannot match any vector in the glass so as to leak any light (red lines in Figure 11c); $k_{spp,air}$ mode on the air/metal interface will match the wave vector in the glass (with certain angle) so as to yield leakage light (the red circle in Figure 11c). Of course the leakage mode happens when the film is thin (around tens of nanometers, so it can leak out). When the film is very thin (smaller than about 25 nanometers), there will be coupling between the two waves which will split the resonant energy further (energy of $k_{spp,air}$ will increase and energy of $k_{spp,glass}$ will decrease). Correspondingly, if the waves are excited with fixed wavelength light, the $k_{spp,air}$ and $k_{spp,glass}$ will shift as well. An optional way to measure the $k_{spp,glass}$ is fabricating periodic structures on the metal/glass interface (period is T), which will match the Bloch periodic condition and shift the effective wave vector with $\frac{2n\pi}{T}$($n$ is integer, shown as the blue circle in Figure 11c). Of course according to the practical experiment design, there are also other configurations. Now let learn some works on LRM.

The early researches about the leakage mode were based on the glass hemisphere which doesn't change the emission angle of the light from the center[13-14], so there is not strict Fourier transform relation but with correct emission angle which is still very useful. The direct usage of LRM in early work was done by Drezet et al[15]. As shown in Figure 12, they used a scanning near field optical microscope (SNOM) tip or a single particle/defect to excite SPPs. The leakage light form two bright arcs on the FBP after the Fourier transformation, which is corresponding to the wave vector of the leakage mode. The radius of the arc also shows the emission angle. The arc indicates that the SPPs propagating on the interface are (2D) spherical divergence waves.

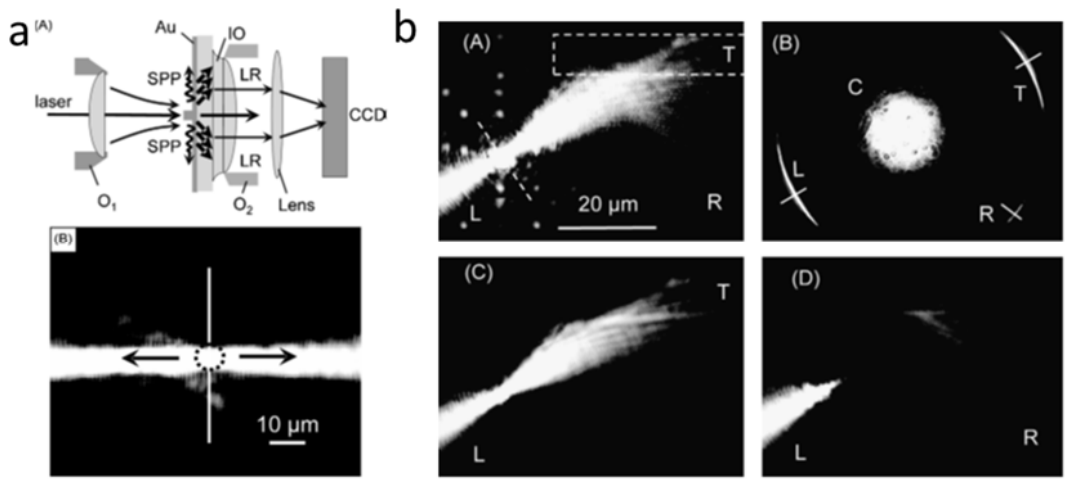

**Figure 12. Measure the propagating direction and wave vector of SPPs with FBP technique. (a) Scheme illustration of the LRM and the leakage mode in real space. (b) Corresponding propagating direction in real space (A, C) and FBP image (B, D). (Figures are copied from ref. 15)[15].**

The typical work done later was performed by Berthelot et al[16]. As shown in Figure 13, they focused the laser beam on the corner of a squared gold film. The defect (the corner) will excite SPPs on both the surface and edges. Because the SPPs on the surface have a divergence angel which is like spherical wave, then there is an arc on FBP. The edge mode SPPs only propagate in y direction, so there is a straight line on FBP perpendicular to $y$ direction shows clear $k_y$ value. It also can be considered having all different wave vectors in $x$ direction because the Fourier transform of a line (a point in x direction) is infinity in frequency (or $k$) space, so it is a line in $x$ direction in FBP with all $k_x$ values. Other similar works please refer to Ref. [17-20].

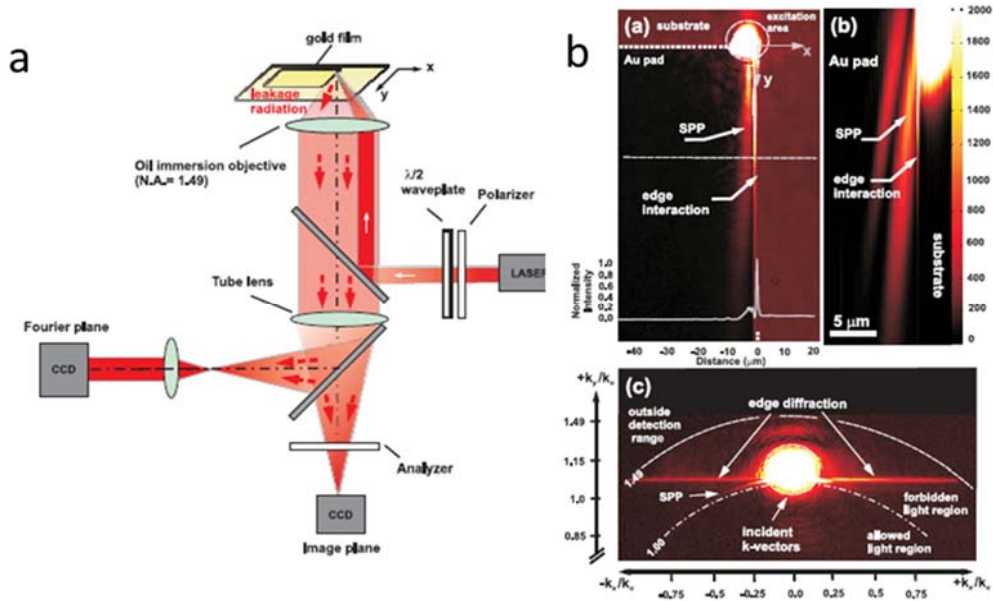

**Figure 13. Image the wave vectors for different SPP modes with FBP. (a) Setup.(b) The images for real space (a, b) and  $k$  space (c) (Figures are reprinted with permission fromRef. 16, OSA)[16].**

Another very typical application is imaging the wave vectors of 1D nano wire waveguide[21]. As shown in Figure 14, for square dielectric waveguides with different width, there are different modes. When exciting with 800 nm light, different modes will be excited in different waveguides (Figure 14a). By using FBP technique, these modes can be imaged (Figure 14b). As shown in Figure 14c, the $TM_{00}$, $TM_{01}$ and $TM_{02}$ modes in the 1.5 $\mu m$ waveguide are excited and they show very clear different positions and pattern on the FBP.

Another example with clearer patterns is shown in Ref. [[22]]. For such periodic wave guide modes, the higher modes are standing waves in the cross section direction which have wave vectors in the direction. Hence, when imaging with FBP, there will be symmetric pattern in the FBP image. Under different analyzing polarization, the modes show very different patterns.

A more general application of the LRM is analyzing the modes and wave vectors for 1D plasmonic wave guides like nano wires. The most obvious feature of the plasmonic waveguides different from dielectric waveguides is that there is no cut diameter for the fundamental mode and the localization of the mode volume is decreasing with the guide diameter decreasing so as that the plasmonic wave guide can break through the diffraction limit to guide light signals. Therefore, the plasmonic wave guides have significant applications in photonic chips. The plasmonic modes of the metal nano wires are closely related to the diameter and surrounding medium of the wire[23]. When the nanowire is put on substrate, the modes on the wire will hybrid between the close modes and yield new modes[24]. These modes can be measured with LRM as well. As shown in Figure 15, the nano wire is about 400 nm in diameter. When it guides light, there is clear wave vector image on the FBP[25]. Other typical applications please refer to Ref. [[26-27]].

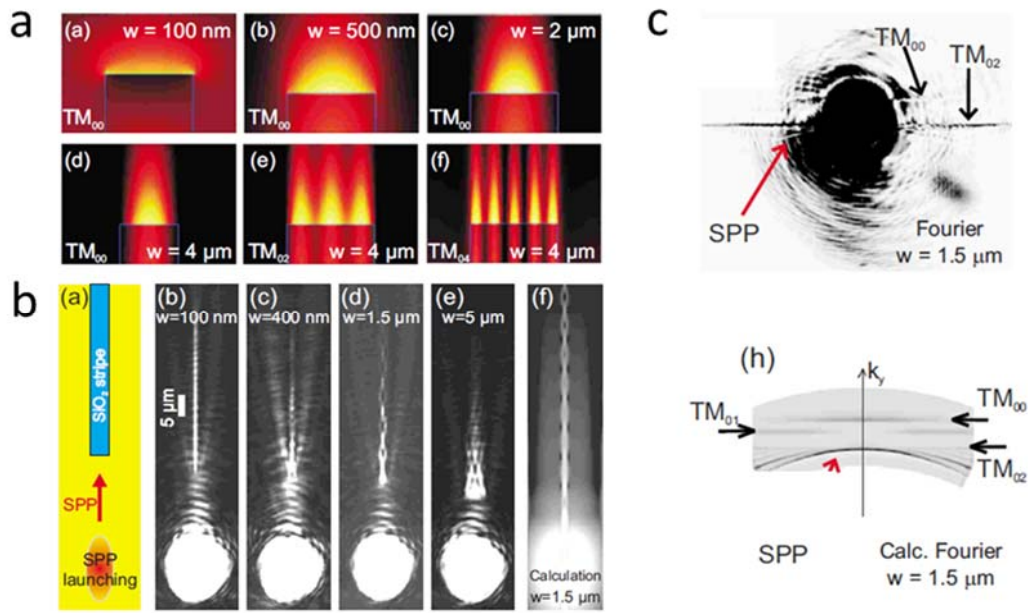

**Figure 14. Image and measure the waveguide modes with FBP. (a) The excited modes of dielectric waveguide with different widths on gold film. (b) The real space image of excited waveguides with different widths by 800 nm laser. (c) The measured and simulated FBP image for the $1.5\ \mu m$ waveguide (Figures are reprinted with permission from ref. 21, APS)[21].**

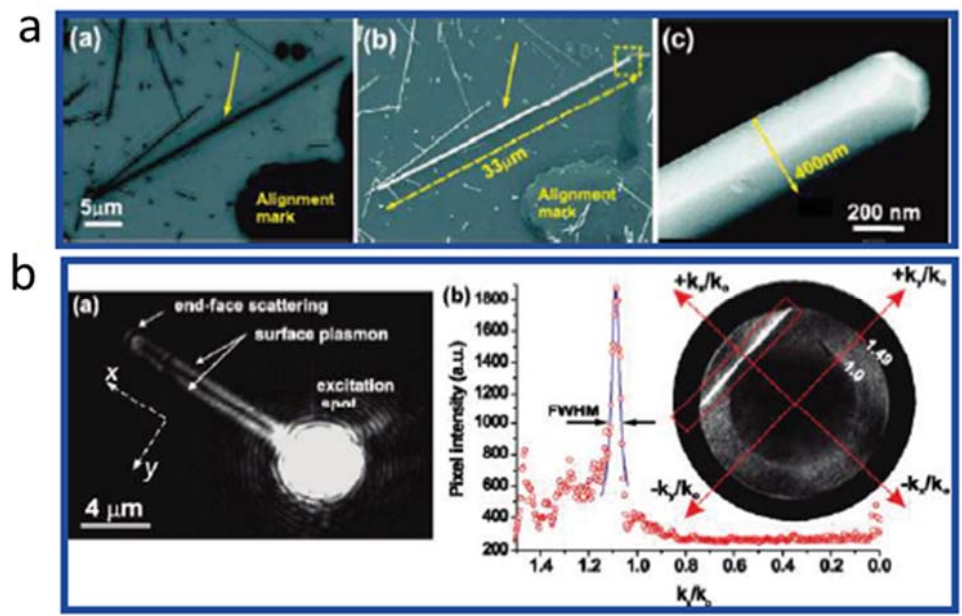

**Figure 15. Measure the silver nano wire waveguide with FBP. (a) The bright field images and SEM images for a nanowire. (b) Real space imaging and FBP imaging for the nano wire waveguide. (Figures are reprinted with permission from ref. 25, ACS)[25].**

### 4.3 Leakage Radiation Microscopy and its applications: dispersion relation measurement

As the LRM can be used to measure the wave vector, it can be used to measure the dispersion relations. Giannattasio et al first measured the dispersion relation by pasting the CCD directly on the substrate, which is still based on the leakage mode principle. In such experiment, the diffractive light from the periodic structures directly detected by CCD (Figure 16a). From the discussion in section 2 we know that such configuration cannot guarantee a strict Fourier transform (and there is an addition uncertain phase factor). The diffraction pattern is received in a very short distance without lens focusing, which is neither a Fraunhofer diffraction nor narrow strip pattern received. The broad pattern will hinder accurate measurement[28].

In section 4.2 we know that the FBP technique can be used to image the wave vectors. Then if the sample is excited with different wavelengths (or white light, or monochromatic excitation with florescence measurement), with the help of prism or gratings, one can directly measure the dispersion relation. As shown in Figure 17, Taminiau et al added a slit to block the light in x direction (so as to let diffractive light with vector in x direction only close to 0 pass) and then image the FBP with an imaging spectroscope. The FBP image (with only y vectors information) is divided into different wavelengths and detected by the CCD. Because of the limit in x direction, the image in x direction shows the wavelength information and vector information in y direction, which is a dispersion relation[29].

Thomas et al measured the dispersion relation of 1D periodic photonic crystal waveguides with different periods by using FBP imaging (Figure 18)[30]. Because there are Bloch modes in the periodic structures which matches different wavelengths, when changing the lattice constant, the resonance modes and wave vectors in the waveguide will change simultaneously. With FBP, the wave vectors of different modes can be directly measured (Figure 18c). The normalized frequency can be calculated with the lattice constants, then the dispersion relations can be drawn (Figure 18d).

The periodic structures without defect (photonic wave guide guiding the light with it defect) dispersion imaging can be just considered as imaging the density of states (DOS) of the 2D crystals [31-32], which will be discussed later.

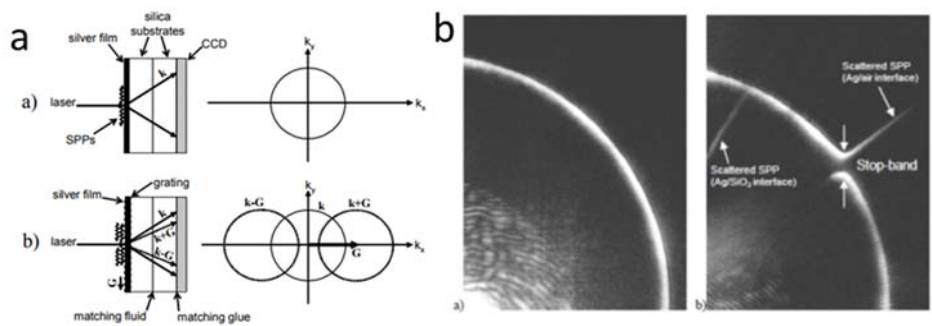

**Figure 16. (a) The setup and schematic illustration for directly imaging the dispersion relation of the leakage modes with CCD and (b) the dispersion images(Figures are reprinted with permission from Ref. 28, OSA)[28].**

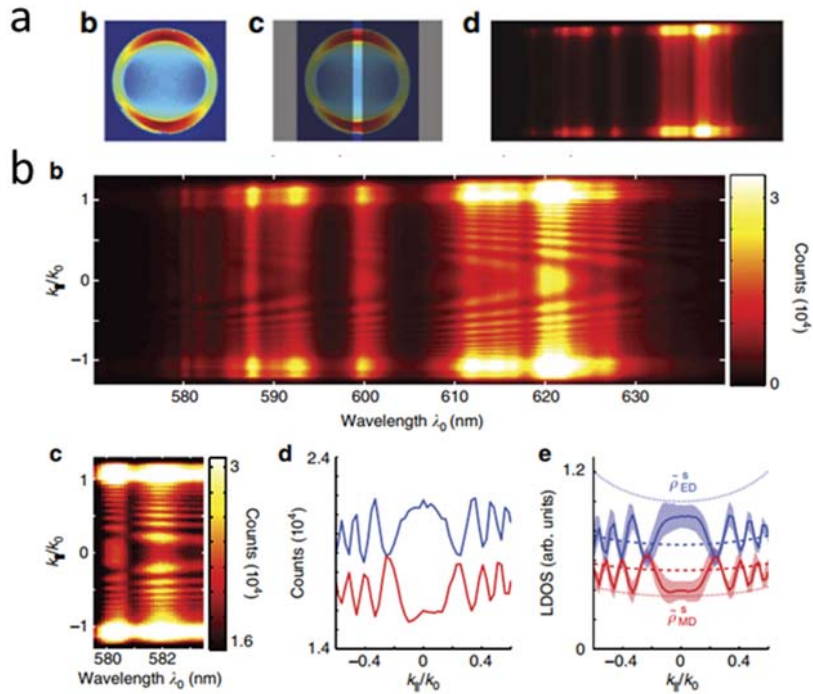

**Figure 17. A method to measure the dispersion relation with FBP technique (Figures are reprinted with permission from Ref. 29, Springer Nature)[29]. (a) There is a slit blocking the rest of part the FBP with only middle part left and then the image is divided into different wavelengths by a spectroscope. (b) The dispersion relation of a sample.**

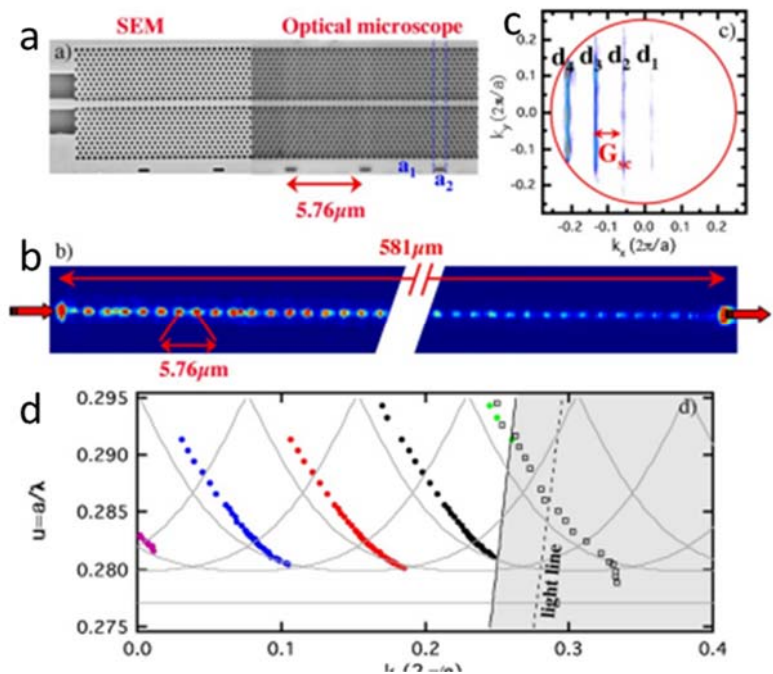

**Figure 18. Measure the dispersion relation of photonic waveguide with FBP technique (a) Photonic waveguide based on GaInAsP/InP. (b, c) The real space and FBP images of the waveguide excited by 1535 nm laser. (d) The dispersion relation curves of the first, second, third and fourth Brillouin zones (Figures are reprinted with permission from Ref. 30, OSA)[30].**

**4.4 Emission angle measurement**

**Principle**

The re-popularity of LRM was originated from the work done by Lieb in 2004. In their work they measured the emission angles of single molecules[1]. As shown in Figure 19, from a direct view, if a light point source is put on the front focus plane, the emission angle distribution can be directly measured on the other side. The early works done in the issue were in the way that the point source was directly put on the flat surface of a hemisphere, because refraction angle on the edge of the hemisphere is 0, the emission angle could be directly read out. However, now we need to investigate such research (more examples like dipole emission or multipole emission of molecules, nano particles and small defect) from the view of Fourier transformation.

Early theoretical works were done by Sommerfeld et al in the beginning of $20^{th}$ century, which were about the emission angular distributions of dipoles parallel or perpendicular to a surface[33-34]. Then between 1977 and 1979 W. Lukosz and R. E. Kunz published three very classical papers, in which they calculated the emission angular distributions of electric dipoles and magnetic dipoles on substrates with very small distances with different orientation in detail[35-37]. In their calculations, even though the dipole is very close to the surface ($z \leq \lambda_1$), they still used the radiation of a dipole in free space first and then taking the far field superposition of reflection and refraction by the interface with Fresnel coefficient. As a matter of fact, the results were very accurate. Later in Lieb's work they performed very almost the same calculations without considering the physics picture underground and got the same results as W. Lukosz's. The main point in their work was analyzing the spatial angular distributions of the molecule emission with FBP technique[1]. The principle of the spatial angular distributions of the molecule emission actually can be analyzed with Fourier transform optics. Just like the model in W. Lukosz's works, when the light radiates from the dipole, it will illuminate the whole surface with different incident angle so as that the parallel components of the wave vectors are different. So he first decomposed the scalar potential of the dipole with Fourier transform, then calculated the Fresnel coefficients of each components[37]. It can understand with the concept of evanescent wave discussed in last section as well. We know that the distance of the dipole to the surface is less than half wavelength, which will certainly excite the evanescent waves on the interface. The wave vectors of the evanescent waves are decided by the boundary conditions, which is related to the polarization and the distance to the surface. According to the reciprocity principle, the evanescent waves will have different leakage modes. So the Fourier transform in such experiment is actually a transformation of the spatial frequency of the evanescent waves. In this way, using the LRM to measure the emission angle of point source agrees with the Fourier transform theory in last sections. *From another point of view,* the Fourier transform of a point on the FBP is corresponding to the whole real space but not just several point in real space. So looking at the position of entrance pupil plane, the light emitted by the dipole is already a result of Fresnel diffraction. After the light pass through the lens, the field on the FBP can be approximately considered as a Fourier transform of the dipole source.

**Experiments**

The most common works done with PBP imaging are the measurement of emission angular distributions of a point source and deduced its orientation. In the beginning work done by Lieb (Figure 19)[1], they used an inverted microscope to build a FBP setup to image a single fluorescent molecule. With the corresponding relation shown in Figure 19b (which is also stated above) one can directly calculated the intensity distribution on FBP according to the emission angle. And conversely, with the intensity distributions on FBP, one can calculate out the emission angular distributions and subsequently get the orientation of the molecule (Figure 19d).

The work done by Harmann et al mentioned in section 4.2 was imaging the SPPs vectors with FBP. Therefrom, if one use the carbon nanotube to excite the SPPs. As the SPPs propagating direction is the same as the nanotube, one could still use the FBP imaged wave vectors to measure the orientation of the carbon nanotubes[27]. Another very typical work on this kind of measurement was done by Curto, in which they measured the angular distributions of multipole modes[38]. As shown in Figure 20, a metal nanorod can support dipole and multipole SPP modes, whose emission spatial angular distributions are like electric dipole, quadruple, octupole and hexadecapole emissions. When it is located on a substrate, very similar as the dipole on substrate above, one can use FBP to image the emission (in this work they excite the SPPs with quantum dot) angular distributions (Figure 20c-n).

Now consider two dipole sources putting on substrate with certain distance, the far field emission pattern will depend on the coherent superposition of the radiation field from the two dipoles[39]. As shown in Figure 21, a gold nano disk (d = 130 nm) and a silver nano disk (d = 110 nm) are put on glass substrate with 15 nm gap. Because the surface plasmon resonant wavelengths are different for the two disks (about 660 nm for the gold disk and 550 nm for the silver disk), they will have different phase shift compared with the incident light phase at specific exciting wavelength. If the wavelength of the incident light is smaller than the resonant wavelength of the plasmon disk, the response of the disk will have a phase difference backward about $\pi$, and contrarily the disk will be the same phase when the incident light's wavelength is larger than the plasmonic resonant wavelength, which is exactly the same as a spring oscillator. So exciting the two disks system with different wavelengths, the interference of the two disks with different responses will emission the light to different spatial angle (Figure 21). With this principle, once can split light into different colors in different angle and then the colors can be measured directly with FBP technique (Figure 21c). We know that the plamon resonance is very sensitive to the surrounding medium refractive index. Therefore, when the environment changes, the shift of the resonance will cause the light color emit to a new angle, which can be used for sensing. In Shegai's work, they directly measured the ratio of light intensity of two colors on FBP for judging if there are molecules adsorbed on the disks (Figure 22)[40-42].

Besides, the FBP technique can be used to directly image the emission angle of light from the terminal of plamonic waveguide. And using the LRM, the wave vectors on the plasmonic waveguide with different modes caused by diameter difference or environment difference can be directly measured[43]. Other similar applications please refer to Ref. [44-45].

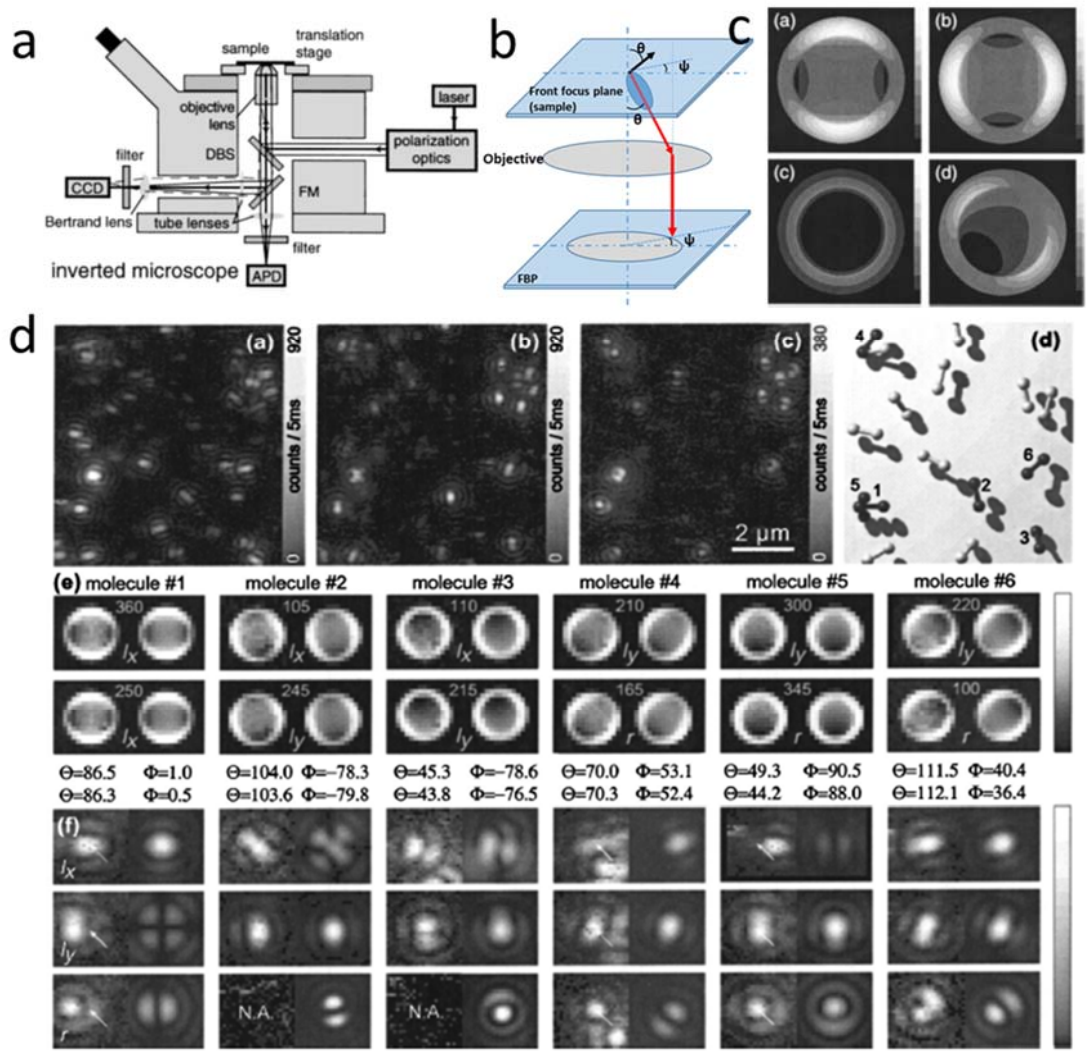

**Figure 19. Image and calculate the orientations of single molecules with FBP technique. (a) Setup. (b) Scheme of emission angle. (c) FBP images for single molecules with different orientations. (d) Real measurement of FBP images for single molecules with different orientations and calculated orientations. (Figure a, c, d are reprinted with permission from Ref. 1, OSA)[1].**

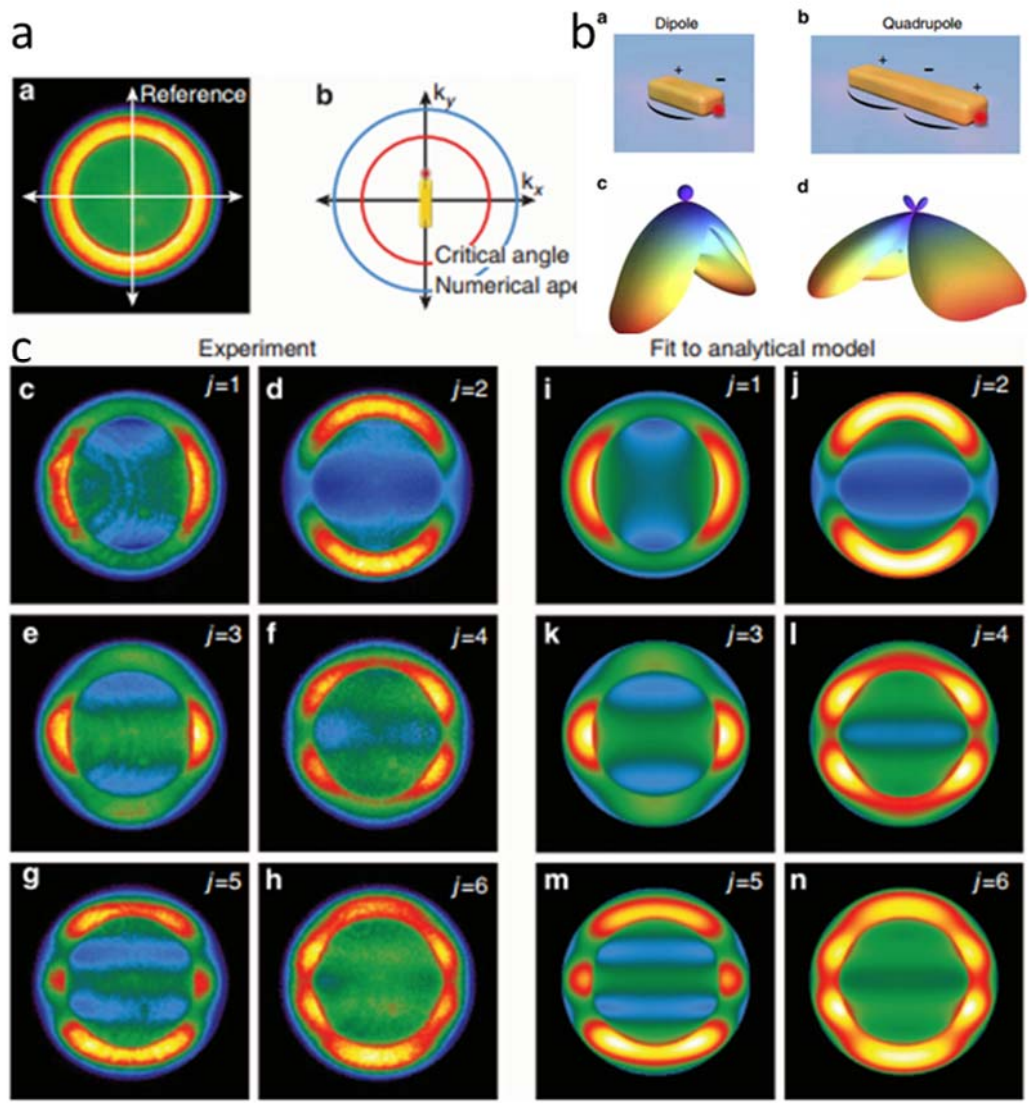

**Figure 20. Image multipoles emission with FBP. (a) Scheme of exciation of a gold nanorod. (b) Simulated emission patterns for dipole and quadrupole modes of gold nanorod. (c) Experimental measurement and simulations of multipole emission angular distributions on FBP (Figures are reprinted with permission from Ref. 38, Springer Nature)[38].**

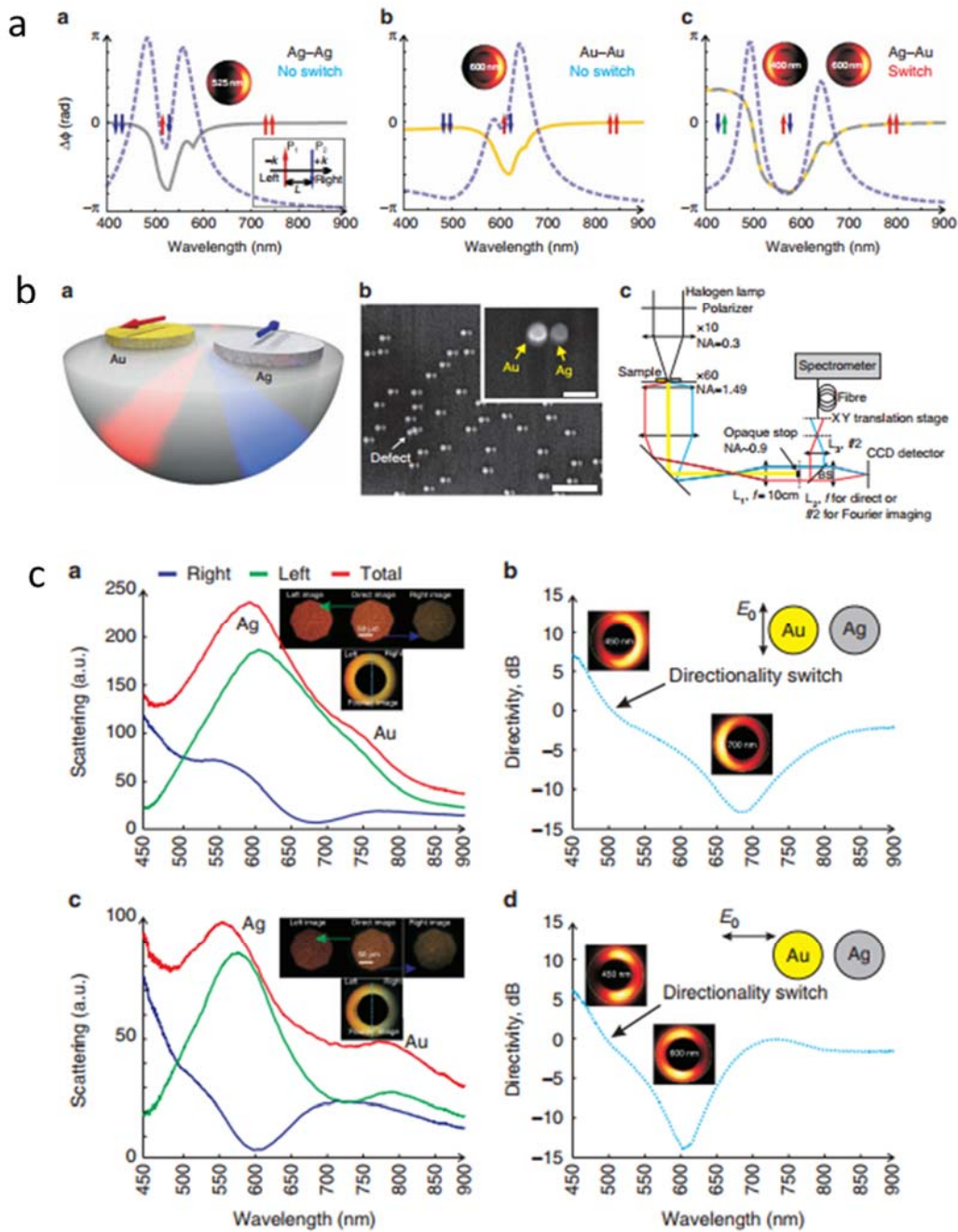

**Figure 21. Measure the spectroscopic phenomenon of two plasmonic disks system with FBP technique. (a) The oscillation phase response curves of disk dimer systems with same material or different materials. (b) Scheme of the spectroscopic phenomenon by disk dimer, samples and setup for the measurement. (c) The spectra measured on two sides of the FBP and the FBP images(Figures are reprinted with permission from Ref. 39, Springer Nature)[39].**

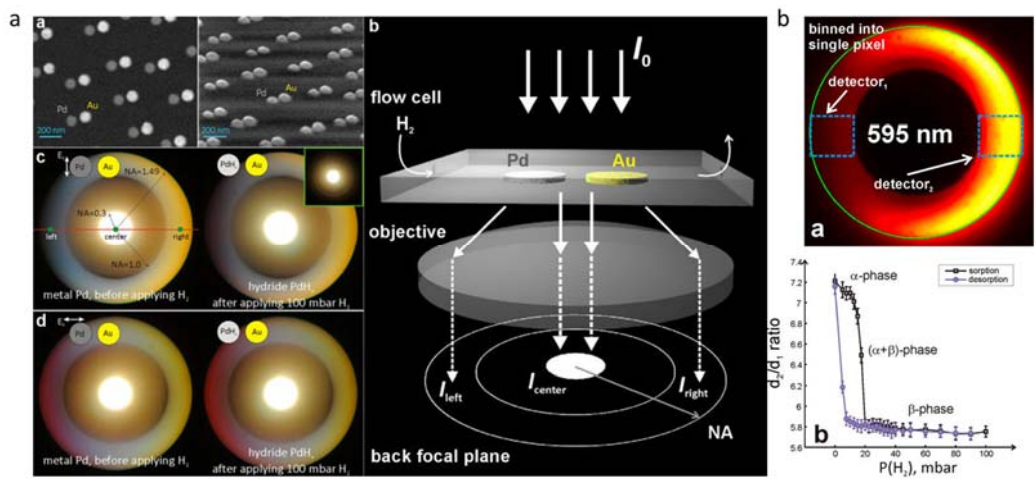

**Figure 22. The sensing experiment by using the FBP and spectroscopic phenomenon of disk dimer described in Figure 21. (a) Sample, FPB scheme and FBP image for the sensing. (b) FBP image measured at 595 nm for $H_2$ adsorbing sensing experiment. (Figures are reprinted with permission from Ref. 40, ACS)[40].**

### 4.5 $k$-space imaging and 2D photonic crystals

The imaging of DOS in $k$ space for the 2D photonic crystals is very similar to the FBP imaging for the wave vectors of waveguides described above. The difference is that the wavevector in photonic crystal is formed by the Bloch periodic boundary conditions, but not the propagating evanescent wave on a planar surface. Intuitively speaking, Fourier transformation transform the information in real space to $k$ space. In the deduction in section 2 we used the spatial frequency. However, we know that because of the Bloch condition, the spatial frequency matches the wavevector in $k$ space. So to image $k$ space is closer to the intuitive picture of Fourier transform in crystals[46-47].

The measurement of dispersion relation in Figure 14 is actually also an example[28]. Now let take a look at the periodic samples in only one direction (Figure 23a)[48]. It is the same and infinity in $x$ direction and periodic in $y$ direction. The periodic structure is made of polymethyl methacrylate (PMMA) rectangular strip waveguides on flat gold film. There are doped rhodamine 6G (R6G) molecules in PMMA for fluorescence imaging. The samples are excited with laser; the light will excite the waveguide modes in the PMMA waveguide. Meantime, there are SPP modes on the interface of PMMA and gold film. The leakage of these mode can be directly imaged on FBP. Another very interesting phenomenon is that there are modes shift in $k_y$ direction because of the

Block wave ($k_{diff,\parallel} = k_{inc,\parallel} + G$, $G = \frac{2\pi}{a}n$, $n$ is integer, Figure 23b). More obvious shift can be observed in the sample shown in Figure 23a(b). In this sample extra Al strips are evaporated on Au film and PMMA strips are after that. Thus, the periodic condition is more abrupt, which causes clearer Bloch wave vector shift.

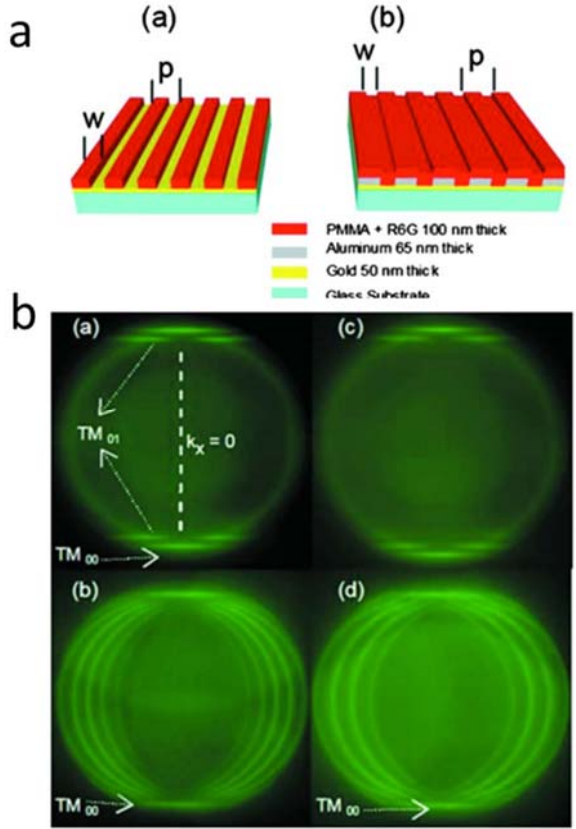

**Figure 23. Imaging the Bloch waves in 1D structures with FBP. (a) Sample schemes. (b) FBP image of sample a(b) under vertical polarization (up row) and level polarization (down row)(Figures are reprinted with permission from Ref. 48, AIP)[48].**

As a Fourier transform setup, one of the advantages is directly image the DOS in $k$ space. As shown in Figure 24[49], there is 50 nm Au film on glass substrate, on which there are 2D PMMA structures doped with R6G. The sample is excited with laser, and the fluorescence of R6G is measured. The light will form SPP Bloch waves in the periodic structure and then leak to the substrate. On FBP now we can directly see the Fermi surface of the photonic structure. As shown in Figure 24b, the inside light color ring represents the SPP wave with an average effective dielectric constant, whose wave vector is decided by the period. The outside ring represents the NA of the objective and the maximum wave vector can be measured. The white square shows the first Brillouin zone (BZ) in $k$ space. Because the wave vectors in other Brillouin zone are shifted $G = \frac{2\pi}{a}n$ because of the translational symmetry. The shift depends on the value of period $a$. If the shifted vector is overlapping with the vector in vector in the First BZ, they will partially cancel each other because they have opposite propagating direction, which makes the strip darker (the dark arcs on the inner ring). From the patterns we can see that they are very similar to the Fermi surface in solid state physics. If we zoom in the patterns, we can see small gap structure (see Figure 16b as well). When the lattice structure changes, the BZ shape changes as well as the Fermi surface (Figure 24 b(c), b(d)). If the lattice constant $a$ changes, the size of the BZ changes as well as shown in Figure

24c.

We know that in solid state physics, the DOS can be calculated from the dispersion relation. When the dispersion curve is flatter at certain frequency, the DOS will increase. A standard way to deduce the DOS is calculate using the periodic boundary conditions in inverse lattice space. But from the previous sections we know that the dispersion relation can be directly measured with FBP. So here if we only image the FBP (DOS distribution in $k$ space) at one frequency each time, then integrate the distributions we can get DOS (for 2D structure with period $L$, the wave vector is $\boldsymbol{k} = n_x \frac{2\pi}{a} \hat{\boldsymbol{x}} + n_y \frac{2\pi}{a} \hat{\boldsymbol{y}}$, in $\boldsymbol{k_x}$ and $\boldsymbol{k_x}$ plane, a mode has volume of $\frac{4\pi^2}{a^2}$). More similar analysis please refer to Ref. [50].

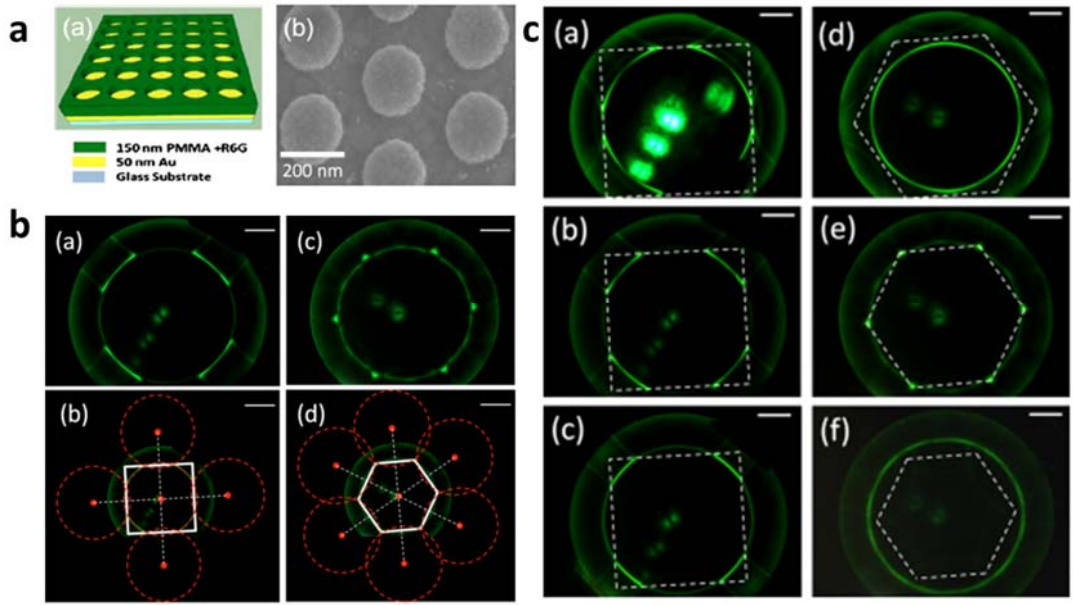

**Figure 24. Image the Fermi surface of 2D photonic crystals with FBP. (a) A sample scheme and a sample SEM image. (b) FBP image for the samples shown in a. (c) FBP images for the samples shown in a but with different period (Figures are reprinted with permission from ref. 49, AIP)[49].**

With the FBP technique, we can not only image the DOS of photonic crystals, but also image the DOS of photonic quasi-crystal structures. It has been known that the crystalline structure has translational symmetry and rotational symmetry (with 1-, 2-, 3-, 4-, 6- fold rotating symmetric axes) in long range, while the quasi-crystalline structure only doesn't have translational symmetry in long rang. Quasi-crystalline structures are kinds of structure that between crystal and amorphous. They have long range order without translational symmetry and have 5-fold or larger than 6-fold rotational symmetric axes. Because of the unique properties, the works on quasi-crystal were rewarded by Nobel price in chemistry in 2011. As Figure 25 shown, the silver nanodisks form Penrose quasi-crystal pattern on glass substrate. The Penrose pattern is consisted with two diamond structures, which doesn't have translational symmetry but with long range order and 5-fold rotational symmetry.

Its Fourier transform pattern shows 10-fold rotational symmetry. These kinds of plasmonic structures don't have periodic near field or far field interaction and their optical extinction property is the same as single nano particles. However, the FBP image shows very obvious ordered structure. As shown in Figure 25b, when imaging the FBP of the structure at different wavelengths, the FBP images show the same 10-fold rotational symmetry as the direction Fourier transform of the structure with computer codes. The diffractive spots of the pattern get further and further from the center as the wavelength increases[51]. The pattern in the middle of Figure 25b is gotten with white light, which is the same as the sum of patterns at different wavelengths. As one knows that the wavevector in plane of the interface satisfies $k_{diff,\parallel} = k_{inc,\parallel} + G$, $G = \frac{2\pi}{a}k$, where $k$ is given by the linear combination of 5-fold rotational symmetry of $e^{i\pi/5}$. $a$ is the period of the first pseudo-Brillion zone. When imaging at the same wavelength, different order of the diffraction patter can be seen clearly when choosing different exposure time. Higher order spots will slowly appear at longer exposure time. These patters can be calculated with the optical displacement-phase shift principle ($I(\theta, \phi) = F(\theta, \phi)S(\theta, \phi) = F(\theta, \phi)\sum_{i,j} e^{iq \cdot r_i}e^{-iq \cdot r_j}$). From Figure 25c we can see that the calculated pattern matches the optical FBP pattern pretty well. More works on FBP imaging photonic quasi-crystals please refer to Ref. [52-53].

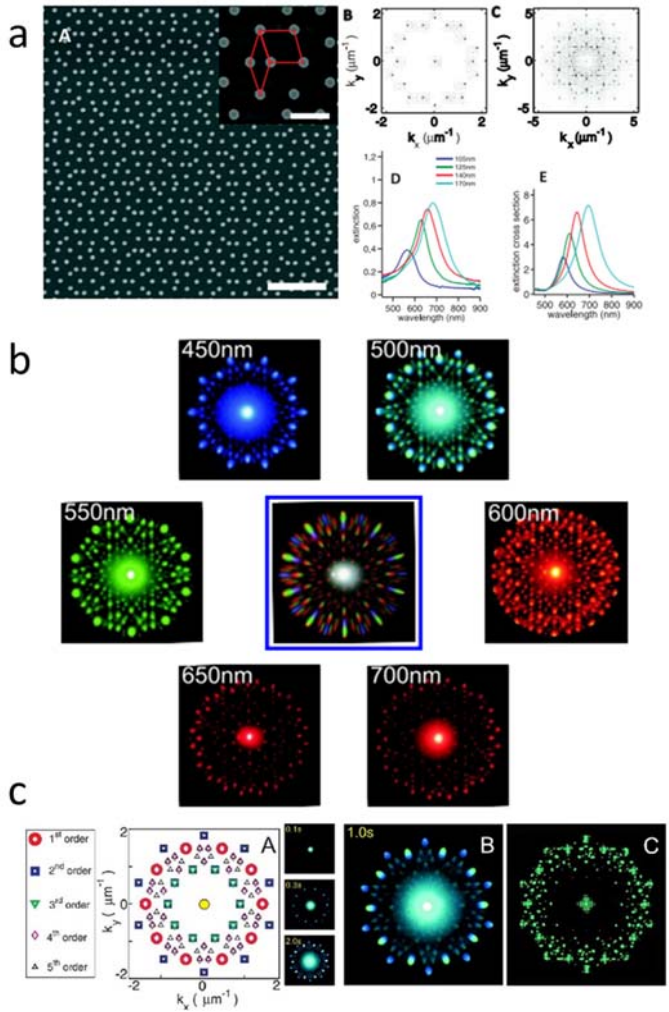

**Figure 25. Image the Brillouin zone of quasi-crystal with FBP. (a) A quasi-crystal sample composed with nano metal disks and its Fourier transform pattern with computer. And the spectra of the samples with different diameters. (b) FBP images of the quasi-crystal under different wavelength filters (white light image is in the middle). (c) FBP pattern of different diffractive order captured by different exploring time. (Figures are reprinted with permission from ref. 51, ACS)[51].**

## 5 Realization in undergraduate teaching lab

Considering that the infinity optical microscope is a little bit expensive and usually there is not much fund for teaching. But there are also a lot of the same cheap microscope in old type in undergraduate student labs. So we need to realize the FBP in the traditional old microscopes.

As shown in Figure 26, one can make an integral tube with Brantrand lenses inside. One side of the tube can be inserted into the eyepiece port which can be purchased through internet and the other end connected to a CCD (also very cheap in web stores). The Brantrand lenses can be switched through a bar with two lenses having different focus lengths. One lens on the switching bar has focus length $f$ (and $2f$ is the distance from the image of the objective and CCD), another lens has focus length $f/2$. Switch the Brantrand lenses will collect the image in real space (focus length $f/2$) or $k$ space (focus length $f$). Because the barrel length of the old type microscope is compatible with the modern infinity optical tube length focus length for the same manufacture, so the small setup can be inserted into either the old type or the modern scientific research microscope. A CD disk can be used as a sample in the student lab. So there is also opportunity in the undergraduate teaching lab to repeat some frontier research setup, which helps the students understand the knowledge in courses and advanced sciences. It could also be an experiment lesson for the students.

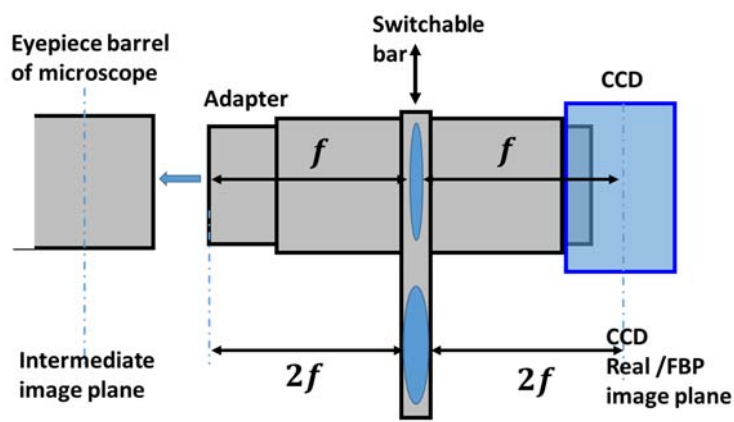

**Figure 26. Schematic design of the FBP setup in a teaching lab.**

## 6 Summary and outlook

In this paper, beginning with Fourier transform frequency division essence of Fresnel diffraction and Fraunhofer diffraction, combing with the approximations in modern infinity optical system, the FBP setups and their applications in frontier research are introduced. With the introduction we can feel how the knowledge in university courses are used in advanced research. In the last a kind of cheap and simple FBP setup is introduced for university experiment.

Fourier transform optics based on the traditional optics has already quite complete since last 60's. It started a new journey in nanophotonics because of the development of nanotechnology. As the developing the nanophotonics and quantum optics, we believe that the Fourier transform imaging technique will continue to push the area. And furthermore, there is no reason to doubt that the Fourier transform optics based on simple lens transmission but with deep complex transformation and operation will play an important role in quantum operations, quantum information processing and parallel photonic chips. If the nano lenses can be fabricated in small scale, which can realize the Fourier transform optics in micro/nano scale, the Fourier technique may make a big step forward.

This paper is based on primary knowledge in university. Hence, the introduction for the advanced research is not in a penetrating way for the studied targets. However, it does not hinder the understanding of the Fourier transform optics part.

**Acknowledgement**

This paper is written in accordance with the guiding principle of transferring advanced scientific research to undergraduate teaching contents in our university. This work was supported by the National Natural Science Foundation of China (No. 11704058) the Fundamental Research Funds for the Central Universities (DUT19RC(3)007).

# 傅里叶后焦面成像原理及应用

方蔚瑞

（大连理工大学 物理学院，大连，116024，*Email: yrfang@dlut.edu.cn*）

## The principle and applications of Fourier back plane imaging

Yurui Fang

*(School of Physics, Dalian University of Technology, Dalian, 116024)*

**Abstract**

Fourier back plane (FBP) imaging technique has been widely used in the frontier research of nanophotonics. In this paper, based on the diffraction theory and wave front transformation principle, the FBP imaging basic principle, the setup realization and the applications in frontier research are introduced. The paper beginnings with the primary knowledge of Fourier optics, combining with the modern microscope structure to help to understand the Fourier transformation principle in the advances of nanophotonics. It can be a reference for experimental teaching and researching.

**Keywords:** diffraction; Fourier back plane; infinity optical system; leakage microscrope; wave vector; k space; photonic crystal

　　**摘要：** 傅里叶后焦面成像近几年来在纳米光学前沿研究中被广泛应用。本文从衍射与波前相位变换的基本原理出发，介绍了傅里叶后焦面成像的基本原理、实验实现、以及近年来在纳米光子学前沿研究的应用。本文从光学基础知识出发，结合现代光学显微镜知识，理解当代相关前沿进展中傅里叶变换的原理及应用，可为研究性实验教学及相关的科研提供一定的参考。

　　关键词：衍射；傅里叶后焦面；无限远光学系统；漏模显微镜；波矢；k 空间；光子晶体

　　中图分类号：O436

## 1 简介

　　傅里叶变换光学是光学这门古老而又新颖的学科里面一个较老而又具有重要当代应用价值的方向。传统的傅里叶变换光学在上世纪六十年代就已经非常成熟。但是随着微纳加工技术的发展、纳米光学的研究以及量子点等小尺寸样品研究的需要，自从 Lieb 等人利用现代显微镜做的工作之后[1]，傅里叶显微镜又在现代技术的基础上重新发展了起来。本文的目的就是基于傅里叶变换光学的基本原理来介绍傅里叶显微技术及其在前沿科技中的应用。

　　我们首先介绍了傅里叶变换光学的原理，然后介绍了在无限远光学系统中傅里叶变换光学的实现。之后利用这些基础知识来分析了傅里叶显微镜在高阶散射高分辨成像、表面波波矢测量、纳米颗粒及小分子的发射方向、光子晶体 k 空间成像等前沿应用。最后我们简单讨论了在大学物理实验室如何简单而又低成本的实现前沿研究中的傅里叶后焦面（Fourier Back plane, FBP）成像显微镜。为了使本文聚焦于基本原理及仪器应用原理，**在前沿研究的介绍中，我们只对现象进行直观的描述而避开了深层理论以免本文失去重心。**

由于我们的被观测物总是远小于物镜的尺寸，因此在以下的讨论中，为了避免陷入复杂的数学处理，我们假定透射窗函数为 1，仅利用我们所熟悉的菲涅尔衍射、夫琅禾费衍射、透镜相位变换以及傅里叶变换的语言来叙述而避免使用角谱理论等。而且在整个过程中只考虑单色光的变换，非单色光近似为不同色光光强的线性叠加。

## 2 傅里叶变换原理 [2]

基本数学原理，函数$U$的二维傅里叶变换的积分表示为

$$F_U(f_x, f_y) = \iint\limits_{-\infty}^{+\infty} U(x, y) e^{-j2\pi(f_x x + f_y y)} dx dy \quad (1)$$

其逆变换为

$$U(x, y) = \iint\limits_{-\infty}^{+\infty} F_U(f_x, f_y) e^{j2\pi(f_x x + f_y y)} df_x df_y \quad (2)$$

傅里叶变换的卷积关系为，如果

$$U_1(x, y) = \iint\limits_{-\infty}^{+\infty} U_2(\xi, \eta) h(x - \xi, y - \eta) d\xi d\xi \equiv U_2(\xi, \eta) \otimes h(x, y) \quad (3)$$

则其傅里叶变换满足

$$F_{U1}(f_x, f_y) = H(f_x, f_y) F_{U2}(f_x, f_y) \quad (4)$$

其中$H(f_x, f_y)$为$h(x, y)$的傅里叶变换

$$H(f_x, f_y) = \iint\limits_{-\infty}^{+\infty} h(x, y) e^{-j2\pi(f_x x + f_y y)} dx dy \quad (5)$$

## 2.1 衍射屏可以看成余弦光栅叠加

由于实际被观测物或者衍射屏为各种原子或分子组成，其大小远小于可见光半波长，由此可以看作是连续的。由此我们假定任何（一维）衍射屏的空间特征都可以用傅里叶变换分解为不同频率的余弦光栅的叠加：

$$t(x) = t_0 + \sum_{n \neq 0} t_n e^{j2\pi f_n x} \quad (6)$$

衍射积分的本质是脉冲响应系统的线性叠加。所以光屏的衍射完全可以用傅里叶变换进行研究。二维衍射屏只是一维情形的直接扩展。

## 2.2 菲涅尔衍射与夫琅禾费衍射作为分频原理

菲涅尔近似和夫琅禾费衍射与傍轴近似是等价的 [2]。

从惠更斯-菲涅尔原理及基尔霍夫积分公式，我们知道一个平面衍射孔屏$U(\xi, \eta)$在场点

（$r_0 \gg \lambda$）处的菲涅尔近似（$z^3 \gg \frac{\pi}{4\lambda}[(x-\xi)^2 + (y-\eta)^2]_{max}^2$）分布为

$$U(x,y) = \frac{e^{jkz}}{j\lambda z} e^{j\frac{k}{2z}(x^2+y^2)} \iint_{-\infty}^{+\infty} [U(\xi,\eta) e^{j\frac{k}{2z}(\xi^2+\eta^2)}] e^{-j\frac{2\pi}{\lambda z}(x\xi+y\eta)} d\xi d\eta \quad (7)$$

由此，菲涅尔衍射的结果是一个与$x^2 + y^2$相关的因子相乘的紧靠孔径屏的复场与一个关于$\xi^2 + \eta^2$二次相位因子的傅里叶变换。由此菲涅尔衍射可以看成是紧邻衍射孔径附近的光场的傅里叶变换。

夫琅禾费近似（$z \gg \frac{k}{2}(\xi^2 + \eta^2)_{max}$）衍射公式为

$$U(x,y) = \frac{e^{jkz}}{j\lambda z} e^{j\frac{k}{2z}(x^2+y^2)} \iint_{-\infty}^{+\infty} U(\xi,\eta) e^{-j\frac{2\pi}{\lambda z}(x\xi+y\eta)} d\xi d\eta \quad (8)$$

相对于菲涅尔衍射，积分里面的因子$e^{j\frac{k}{2z}(\xi^2+\eta^2)}$消失了。可以看出夫琅禾费衍射除了与$x^2 + y^2$相关的因子二次方的相位因子，就是衍射孔径上场分布的傅里叶变换，变换频率为

$$f_x = x/\lambda z \quad (9a)$$
$$f_y = y/\lambda z \quad (9b)$$

由于夫琅禾费近似所需要的条件非常苛刻，一般我们利用增加透镜的方式在实验室实现。所以说利用透镜实现的夫琅禾费衍射本身就是一个傅里叶变换。夫琅禾费衍射可实现衍射分频。

## 2.3 薄透镜变换函数

考虑如图 1 所示的一般透镜的相位延迟

$$\phi(x,y) = kn\Delta(x,y) + k[\Delta_0 - \Delta(x,y)] \quad (10)$$
$$t_l(x,y) = exp[jk\Delta_0] exp[jk(n-1)\Delta(x,y)] \quad (11)$$

$$\Delta(x,y) = \Delta_0 - R_1 \left(1 - \sqrt{1 - \frac{x^2+y^2}{R_1^2}}\right) + R_2 \left(1 - \sqrt{1 - \frac{x^2+y^2}{R_2^2}}\right) \quad (12)$$

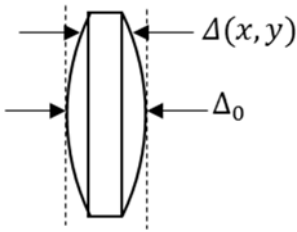

**图 1 透镜光程函数计算示意图，其中左侧球面半径为$R_1$，右侧球面半径为$R_2$，当光从左侧入射，则$R_1$取正，$R_2$取负。**

由于考虑光强，我们忽略常相位因子$\Delta_0$

傍轴条件下

$$\sqrt{1 - \frac{x^2 + y^2}{R_i^2}} \approx 1 - \frac{x^2 + y^2}{2R_i^2} \quad (13)$$

$$\Delta(x,y) = \Delta_0 - \frac{x^2 + y^2}{2}\left(\frac{1}{R_1} - \frac{1}{R_2}\right) \quad (14)$$

$$\frac{1}{f} = (n-1)\left(\frac{1}{R_1} - \frac{1}{R_2}\right) \quad (15)$$

$$t_l(x,y) = exp[jk(n-1)\Delta(x,y)] = exp\left[-jk\frac{x^2 + y^2}{2f}\right] \quad (16)$$

如果平面波入射且场分布为 1，则透射后场分布为

$$U_l(x,y) = exp\left[-j\frac{k}{2f}(x^2 + y^2)\right] \quad (17)$$

为球面波。

## 2.4 薄透镜的傅里叶变换性质

考虑光路如图 2a 所示，输入物屏（透射函数为$t_o(x,y)$）平行于透镜平面紧挨着透镜放在透镜前面，用垂直于屏的振幅为$A$的平面波照射，则透射后振幅分布为

$$U_l(x,y) = At_o t_l = At_o \, exp\left[-jk\frac{x^2 + y^2}{2f}\right] \quad (18)$$

利用菲涅尔衍射公式，在透镜的后焦面（$z = f$）上，忽略定常相位因子$e^{jkf}$，我们有

$$U_f(u,v) = \frac{e^{j\frac{k}{2f}(u^2+v^2)}}{j\lambda f}\iint\limits_{-\infty}^{+\infty} U_l(x,y)e^{j\frac{k}{2f}(x^2+y^2)}e^{-j\frac{2\pi}{\lambda f}(xu+yv)}dxdy$$

$$= \frac{e^{j\frac{k}{2f}(u^2+v^2)}}{j\lambda f}\iint\limits_{-\infty}^{+\infty} At_o(x,y)e^{-j\frac{2\pi}{\lambda f}(xu+yv)}dxdy \quad (19)$$

我们可以看出，这时的光场分布与夫琅禾费衍射公式 8 完全一致。这也是由于在透镜存在的情况下，后焦面上的成像可以看成是入射方平行光线在无穷远处成像，这样是满足夫琅禾费衍射条件的，这也是我们在实验室中利用透镜来实现夫琅禾费衍射的原因。在这种情况下，由于积分号前面的二次相位因子的存在，这个傅里叶变换关系还不是确定的。

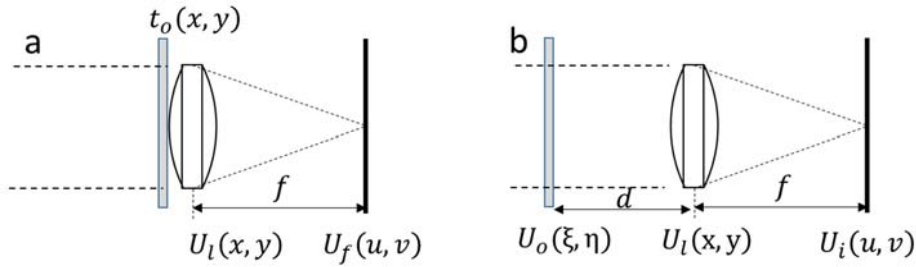

**图 2** 正透镜后焦面成像光路图。

接下来我们考虑一般光路如图 2b 所示，输入物屏（透射函数为$t_o(\xi,\eta)$）平行于透镜平面放在透镜前面距离d处，用垂直于屏的振幅为 A 的平面波照射。为了方便以下的推导，我们令$F_o(f_\xi, f_\eta)$为输入衍射物屏光场$U_o = At_o(x,y)$的傅里叶频谱，$F_l(f_x, f_y)$为物屏的光信号

投射到透镜入射面的位置的光场 $U_l(x, y)$ 的傅里叶频谱。则我们可以写出投射到透镜后焦面（$z = f$）上的光场为（忽略定常相位因子 $e^{jkf}$）

$$U_i(u, v) = \frac{e^{j\frac{k}{2f}(u^2+v^2)}}{j\lambda f} \iint\limits_{-\infty}^{+\infty} U_l(x, y) e^{-j\frac{2\pi}{\lambda f}(xu+yv)} dxdy = \frac{e^{j\frac{k}{2f}(u^2+v^2)}}{j\lambda f} F_l(f_x, f_y) \ (20)$$

其中 $f_x = \frac{u}{\lambda f}, f_y = \frac{v}{\lambda f}$

由菲涅尔衍射公式我们知道

$$U_l(x, y) = \frac{e^{jkz}}{j\lambda z} e^{j\frac{k}{2z}(x^2+y^2)} \iint\limits_{-\infty}^{+\infty} [U_o(\xi, \eta) e^{j\frac{k}{2z}(\xi^2+\eta^2)}] e^{-j\frac{2\pi}{\lambda z}(x\xi+y\eta)} d\xi d\eta$$

$$= \iint\limits_{-\infty}^{+\infty} U_o(\xi, \eta)[\frac{e^{jkd}}{j\lambda d} e^{j\frac{k}{2d}((x-\xi)^2+(y-\eta)^2)}] d\xi d\eta = U_o(\xi, \eta) \otimes h(x, y) \ (21)$$

其中 $h(x, y) = \frac{e^{jkd}}{j\lambda d} e^{j\frac{k}{2d}(x^2+y^2)}$，其傅里叶变换为

$$H(f_x, f_y) = \iint\limits_{-\infty}^{+\infty} h(x, y) e^{-j2\pi(f_x x + f_y y)} dxdy = e^{jkd} e^{-j\pi\lambda d(f_x^2+f_y^2)} \ (22)$$

其中 $f_x = \frac{u}{\lambda d}, f_y = \frac{v}{\lambda d}$，所以

$$F_l(f_x, f_y) = H(f_x, f_y) F_o(f_\xi, f_\eta) \ (23)$$

代入式 20，我们有

$$U_i(u, v) = \frac{e^{j\frac{k}{2f}(u^2+v^2)}}{j\lambda f} H(f_x, f_y) F_o(f_\xi, f_\eta) = \frac{e^{j\frac{k}{2f}(u^2+v^2)}}{j\lambda f} e^{jkd} e^{-j\pi\lambda d(f_x^2+f_y^2)} F_o(f_\xi, f_\eta)$$

$$= g(u, v) F_o(f_\xi, f_\eta) \ (24)$$

其中

$$g(u, v) = \frac{e^{j\frac{k}{2f}(1-\frac{d}{f})(u^2+v^2)}}{j\lambda f} \ (25)$$

在运算中我们忽略了常相位因子 $e^{jk(d+f)}$.

由式 24 我们可以看出，在透镜后焦面上的光场是物屏 $U_o(\xi, \eta)$ 的傅里叶变换与一个二次项因子 $g(u, v)$ 的乘积。当 $d = 0$ 时，式 24 回归为夫琅禾费衍射公式 8 及物屏紧邻薄透镜的公式 19。当 $d = f$ 时，$g(u, v)$ 变为一个复常数数矩，不影响后焦面的光场分布，此时透镜后焦面上的光场完全是前焦面上物屏场 $U_o(\xi, \eta)$ 的傅里叶变换。当 $d = \infty$ 时，我们同时考虑常相位因子 $e^{jk(d+f)}$，若 $u^2 + v^2 \ll f^2$，$g(u, v) \approx \frac{e^{jkd}}{j\lambda f}$ 也是一个常相位因子，所以这时也满足傅里叶变换关系，我们稍后将用到这个结论。

## 2.5 例子：频谱分析、余弦光栅

从 2.1 节我们知道，余弦光栅可以作为最基本的衍射单元来分解各种不同的衍射屏。我们下面利用一维余弦光栅作为衍射屏来体会一下上面的傅里叶变换过程。假设余弦光栅的透过率函数为

$$t_{cos}(x,y) = t_0 + t_1 cos(2\pi f_x x + \psi_0) \quad (26)$$

当一束单色平行光（波前为$U_0$）照射在这个光栅上面，假设$\psi_0 = 0$，经过余弦光栅之后的波前为

$$U_{cos}(x,y) = U_0 t_{cos}(x,y) = U_0(t_0 + t_1 cos(2\pi f_x x))$$

$$= U_0 t_0 + \frac{1}{2} U_0 t_1 e^{j(2\pi f_x x)} + \frac{1}{2} U_0 t_1 e^{-j(2\pi f_x x)}$$

$$= U_0 t_0 + \frac{1}{2} U_0 t_1 e^{jk(f_x\lambda x)} + \frac{1}{2} U_0 t_1 e^{-jk(f_x\lambda x)}$$

$$= U_0 t_0 + \frac{1}{2} U_0 t_1 e^{jk_x x} + \frac{1}{2} U_0 t_1 e^{-jk_x x} = \widetilde{U}_0 + \widetilde{U}_1 + \widetilde{U}_{-1} \quad (27)$$

如果此余弦光栅位于透镜前焦面上，则透镜后焦面上的光场为（与y方向无关）

$$U_i(u,v) = \frac{1}{j\lambda f} F_o(f_\xi, f_\eta) = \iint\limits_{-\infty}^{+\infty} (\widetilde{U}_0 + \widetilde{U}_1 + \widetilde{U}_{-1}) e^{-j2\pi(f_x x + f_y y)} dx dy$$

$$= 2\pi\delta(u-0)\delta(v-0) + \pi\delta(u+k_x)\delta(v-0) + \pi\delta(u-k_x)\delta(v-0) \quad (28)$$

可以看出一维余弦光栅在薄透镜焦面上是三个亮点，其中中间一点位于光轴上，另外两点位于两侧并且与第一点对称分布（图3）。其中$k_x = 2\pi f_x = 2\pi/d_x$，反映了物屏的周期以及光波在$x$方向的波矢，$k_y = 0$。

对于其他周期性函数，我们可以看成是多个不同周期的余弦光栅叠加而成，每个余弦光栅在透镜后焦面上的光场分布都具有相应的频率点，由此达到分频的目的。

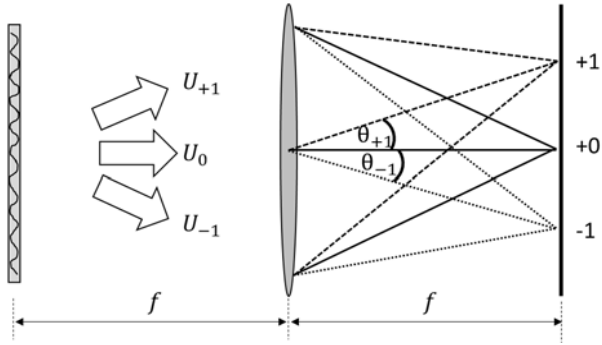

**图3 单色平面波正入射照明时余弦光栅1级衍射示意图。**

## 2.6 从衍射、薄透镜傅里叶变换的角度看衍射极限与近场光学 [3]

由式 27 的表达式我们可以看出，$\widetilde{U}_0$、$\widetilde{U}_1$、$\widetilde{U}_{-1}$各代表一列平面波（图4a）。$\widetilde{U}_0$代表一列正出射的平面波，我们成为 0 级波，其在$x$方向上没有波矢分量。$\widetilde{U}_1$代表一列斜向上出射的平面波，$\widetilde{U}_{-1}$代表一列斜向下出射的平面波，其倾斜角满足

$$sin\theta_{\pm 1} = \pm f_x\lambda = \pm\frac{\lambda}{d_x} = \frac{\pm n\lambda}{n d_x}，\ n = 1,2,3,\dots \quad (29)$$

这三列平面波经过透镜聚焦以后分别汇聚于三个焦点，即余弦光栅的衍射斑。公式 29

正好是夫琅禾费一级衍射公式，或者看成是周期为 $nd_x$ 的光栅的高阶衍射。

令 $f_0 = 1/\lambda$ 代表载波的空间频率，则我们可以将物屏上的空间结构分为三级：$f_x \ll f_0$（低频结构），$f_x \le f_0$（高频结构），$f_x > f_0$（超精细结构）。我们考虑 $f_x > f_0$，则此时 $sin\,\theta_{\pm 1} = \pm f_x\lambda = \pm\frac{f_x}{f_0} > 1$，这时解出的 $\theta_{\pm 1}$ 为复数，此时的衍射场由传播场变为衰减的隐失场（图 4b）。为了更明确看到这一点，我们写出沿此方向传播的平面波的表达式：

$$U_{+1} = U_0 e^{j(k_x x + k_y y + k_z z)} \quad (30)$$

其中

$$k_x^2 + k_y^2 + k_z^2 = k^2 = (2\pi/\lambda)^2 = (2\pi f_0)^2 \quad (31)$$

同时，在衍射光栅出射面有

$$k_x = 2\pi f_x, \ k_y = 0 \quad (32)$$

所以

$$k_z = \sqrt{k^2 - (k_x^2 + k_y^2)} = 2\pi\sqrt{f_0^2 - f_x^2} = 2\pi f_0\sqrt{1 - (\frac{f_x}{f_0})^2} = jk_z' \quad (33)$$

其中 $k_z' = 2\pi f_0\sqrt{(\frac{f_x}{f_0})^2 - 1}$ 是实数。这种波沿着 z 方向急剧衰减，被称为隐失波。这也意味着超精细结构的信息无法通过透镜传递到远场从而被成像。所以从这个方面来看一般显微镜的分辨率极限，也只能是波长量级。

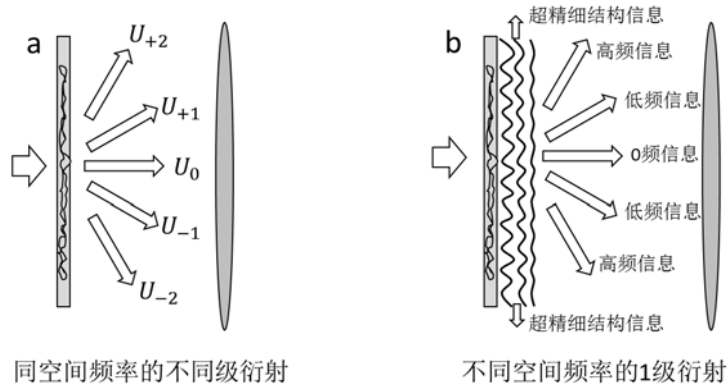

图 4 单色平面波正入射时衍射屏衍射分频示意图。

在透镜的成像过程中，一般来说，由于物镜的口径有限，物体结构的高阶衍射也不能被收集到，所以很自然的会丢失一些信息。有人可能认为，可以通过增加物镜的口径而趋近于无穷大，从而可以突破衍射极限。但是从上面的讨论我们知道，增大物镜口径，只能收集到更多的高阶衍射的光，以及更加逼近波长尺度的结构的衍射光，并不能（收集隐失场信息而）突破衍射极限。通过增加物镜与样品之间空间的折射率（如油浸入式物镜）可以减小衍射角 $\theta_{\pm 1}$ 从而增大数值孔径 $NA = nsin\theta_{\pm 1}$（变相增大 $f_x$）让更加精细结构的光被收集到，也可以增加分辨率。但是一般匹配的油的折射率无法增加太多，所以从傅里叶光学的角度来讲，衍射极限也无法突破。所以目前非扫描式光学显微镜增加分辨率（我们在此不讨论基于扫描式及点扩散函数的数学处理来实现超分辨的显微镜）的主要途径就是减小波长 $\lambda$（增大 $f_0$）的

方式。其中利用短波成像以及高次谐波来实现超分辨是比较常用的手段。所以由于超精细结构的衍射场为隐失场无法传播到远场，我们只能利用近场光学的手段来准确探测超精细结构（基于点扩散函数的扫描式远场超分辨显微镜也只能给出远场信息）。

## 3  傅里叶后焦面(FBP)成像装置

除去传统的利用透镜组的傅里叶变换光学，近年来傅里叶后焦面成像在表面等离激元光子学以及光子晶体领域重新被科学家所利用。但是由于表面等离激元光子学领域的重新兴起是由于微纳加工技术所带来的。而对于微纳样品的测量不可避免的会用到显微镜来进行有效的观察。我们接下来首先看看在现代显微镜系统中傅里叶后焦面成像原理及装置。

### 3.1  无限远光学系统满足夫琅禾费衍射条件及近似薄透镜条件

**无限远光学系统**

现在科研用显微镜经过不断的改进，早已区别于早期的显微镜。早期的显微镜（图 5）样品放在物镜焦距与 2 倍焦距之间，在镜筒中形成倒立放大的实像，然后利用 CCD 对此实像直接成像，或者利用目镜进行观察。其中放大倍数为物镜放大倍数乘以目镜（或者 CCD）放大倍数。在这种设计中，镜筒的长度是一定的，不同的厂家具有不同的标准，一般为 160 mm，180 mm 或者 200 mm，所以缺点也很明显，就是如果要想加入偏振片等其他光学元件，整个镜筒的长度就得重新设计，非常麻烦。现在科研用显微镜为无限远光学系统（图 5）。在无限远光学系统成像过程中，样品位于物镜的焦面上，经过物镜以后，样品上每一点的发出的光线都变成了平行光，然后利用结像透镜（Tube lens）聚焦后在结像透镜的后焦面上形成倒立的实像，之后利用目镜或者 CCD 进行观察。在此过程中，由于经过物镜以后光线为平行光，所以可以很自由的在中间增加各种两面平行的光学元件而不改变光路。由于是平行光，原则上镜筒可以无限延伸，所以也被称为无限远光学系统。而结像透镜的焦距是固定的，各个厂家都与自己的老式设计标准相一致（160 mm，180 mm 或者 200 mm）。而实像的放大倍数为结像透镜的焦距除以物镜的焦距。

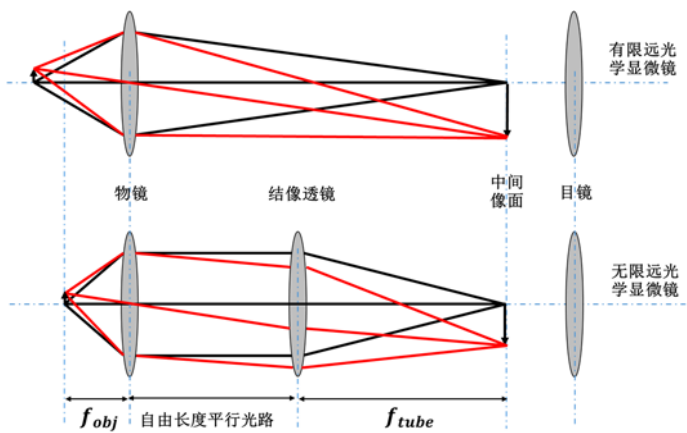

图 5  有限远与无限远光学显微镜示意图。

**无限远光学系统中的物镜及结像透镜**

无限远光学系统实际所采用的物镜（图 6）具有平场消球差、复消色差、消像差、彗差

等各种设计及其组合设计，精巧的设计使得现代物镜都满足傍轴条件。目前没有查到文献直接利用傅里叶光学的分析方法对薄透镜近似以及在这种多透镜组成的复杂物镜中傅里叶变换进行分析，这种分析可能只存在与厂商的设计之中。从专利图（图 6b）的光线路径看，如果将物镜看为一个黑盒子，只考虑入瞳与出瞳，物镜可以满足薄透镜近似条件。尽管在厂商的专利中也没给出所有的参数，但是 Kurvits 等人还是根据专利中的参数，并且结合实际拆开的物镜，对各种品牌及各种规格的物镜用几何光学方法进行了研究（图 7）[4]。研究表明，部分物镜在数值孔径（$NA$）大于 1.3 时，会出现比较严重的畸变。而且高的 $NA/n_{immerge}$ 会导致较高的畸变。一般来说，需要选择低放大倍数及高数值孔径的物镜。对于一定的数值孔径，较小放大倍数将能得到较大的傅里叶后焦面图像，而且图像并不会随着焦距的增大而变模糊。在实际的选择中，由于我们需要一定的放大倍数，所以我们只能选择较高折射率的物镜油，以及较大的物镜出瞳孔径。在后方搭建傅里叶面的地方，选择较长的焦距的透镜。而随着发光物点尺寸的减小，这种畸变会越来越小。

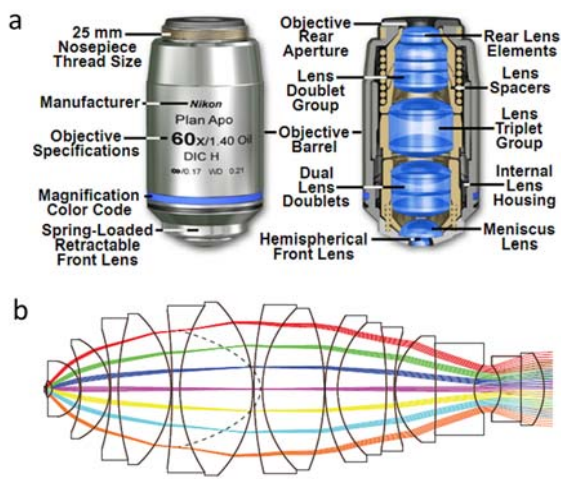

图 6 (a)一种无限远消球差消色差平场油浸入式显微物镜。（Nosepiece Thread Size：物镜接头螺纹尺寸；Manufacturer：制造商；Objective Specifications：物镜参数；Magnification Color Code：放大倍数色环；Spring-Loaded Retractable Front Lens：弹性可缩回前透镜；Objective Rear Aperture：物镜后出光孔；Lens Doublet Group：双胶合透镜组；Objective Barrel：物镜镜筒；Dual Lens Doublets：二重双胶合透镜组；Hemispherical Front Lens：半球形前透镜；Rear Lens Elements：后透镜元件；Lens Spacers：透镜垫；Lens Triplet Group：三胶合透镜组；Internal Lens Housing：中间镜腔。Meniscus Lens：半月透镜。）（图片许可转载自参考文献 5，Nikon）[5]。(b) 一种无限远物镜的内部光路图(图片复制自参考文献 6)[6]。

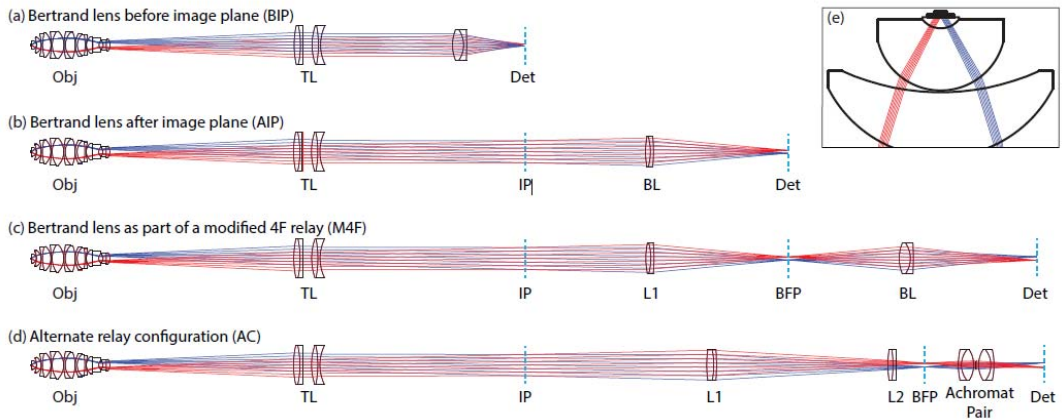

图 7 利用几何光学方法对无限远光学显微镜傅里叶后焦面成像畸变程度的研究。（a）伯特朗透镜置于像面前方的配置。（b）伯特朗透镜置于像面后方的配置。（c）伯特朗透镜用于 4F 系统的配置。（d）另一种可选择配置。（Obj：物镜；TL：结像透镜；Det：探测器；IP：像面；BL：伯特朗透镜；L1：透镜 1；L2：透镜 2；BFP：傅里叶后焦面；Achromat Pair：消色差透镜组。）（e）物镜牵头经光路示意图。（图片许可转载自参考文献 4，OSA）[4]。

　　另外在科研中一般为了能够直接测量样品的散射角度等，我们一般利用高倍（>50x）的油浸入式物镜，这种物镜一般都是平场复消色差物镜。尽管 Kurvits 等人的研究[4]表明对于低倍的平场物镜，其傅里叶变换效果最好，但是在实际高倍测量中，我们只关注视场中心很小的区域的信号，所以在这些测量中都满足我们以上讨论的近似条件。

**照明与成像**

　　明场为科勒照明系统，平行光入射，暗场为直接发射，满足惠更斯-菲涅尔二次波源的条件，所以可以直接利用菲涅尔衍射条件，从而也满足之前的 7 式（从另外一个方面来看，暗场照明方式激发的散射光可以看成是球面波照明，此时如果物屏位于透镜前焦面，在后焦面上的场与物屏场也满足严格的傅里叶变换关系）。

　　在显微测量中，无限远光学系统的样品是位于物镜焦点上的，所以一定满足之前严格傅里叶变换条件。但是，由于物镜的共轭焦点位于物镜镜筒内部，我们无法在物镜后焦面上直接进行傅里叶成像测量，一般的做法是先经过结像透镜在结像透镜后焦面（一般在安装成像 CCD 的接口处，此焦面在显微镜外部）上成一个倒立的实像，然后在显微镜外部利用薄透镜对此倒立放大的实像进行傅里叶变换。

### 3.2 傅里叶后焦面成像装置

　　在实际的科研中，将傅里叶变换透镜波特兰镜（不改变原有光路共轭焦面位置）置于光路不同的部分，傅里叶后焦面成像装置具有四种不同的搭建方式（图 7）[4]。但是一般为了在测量中保证测量样品或者样品部分的准确性，我们一般需要同时采集显微图像。同时考虑到方便性，后焦面的放大倍数等因素，一个比较方便的做法是利用4f系统（尽管 Kurvits 等

人的分析显示 BIP 及 AIP 配置会稍好一些，但是一般在研究中会考虑实际的实现问题），通过转动改变最后一个透镜的焦距从而实现采集实像还是傅里叶变换。如图 8 所示的一个4$f$系统，透镜 1 与透镜 2 是具有相同焦距的薄透镜，一般不镀膜。结像透镜成的一个实像位于 3 处，此像位于透镜 1 的前焦面，同时此像也位于结像透镜的后焦面上，所以可以看成是一个球面波照明。经过透镜 1 之后，在后焦面上将成严格的傅里叶变换，此时如果将透镜 2 的前焦面与透镜 1 的后焦面重合，则在透镜 2 的后焦面上成一个正立的实像，一般我们将 CCD 安置在透镜 2 的后焦面上。而如果我们利用一个光学转轮将透镜 2 替换为透镜 2′，其焦距为透镜 2 焦距的一半，则此时在 CCD 上将成透镜 1 后焦面的倒立等大的实像，亦即像面 3 所成像的傅里叶变换像。

在这个显微镜外的成像过程中我们可以看出对于单色光来说，是满足严格的傅里叶变换关系的。在实际的实验中，对于白光成像的傅里叶变换来说，由于勃兰特透镜的并不能在全波段进行色散校正，傅里叶变换会呈现出一定的偏差，但是在实际科研中，由于选择的透镜焦距的标定位于研究光源的中心波段，所以一般认为是满足傅里叶变换关系的，不做进一步的讨论；而且实际实验中一般利用焦距很大的透镜来搭建装置，这样也能减小色散的影响。

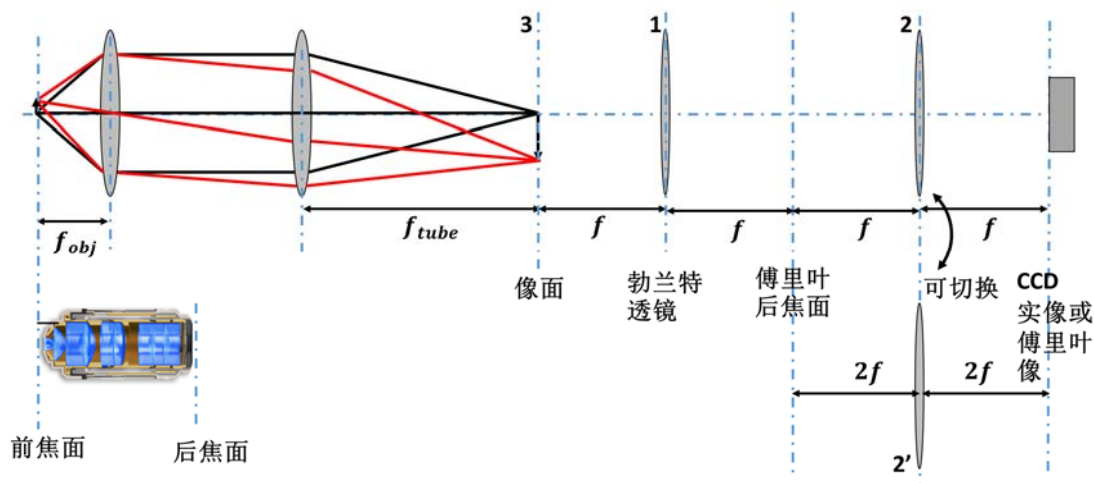

**图 8 傅里叶后焦面成像装置示意图，左下角物镜来自于尼康网站(图片许可转载自参考文献 5，Nikon)[5]。**

另外一个方面就是此傅里叶变换是较严格的针对结像透镜所成的像的一个傅里叶变换，而被研究样品是位于物镜前焦面的实际物体，在这个过程中装置是否满足了结像透镜的像严格反映了待测物的性质。从上一小结的讨论中我们看出在物镜部分，傍轴条件与薄透镜近似条件都是满足的。在无限远光学系统中，结像透镜一般是两个透镜直接胶合而成，而且在无限远光学系统中，结像透镜的入射角很小，满足傍轴条件。而在结像透镜成像时，由于物镜的出射光是平行光（针对某一点来说，满足点扩散函数条件及线性响应条件），所以在无限远光学系统的镜筒中，可以看成是物镜后焦面的二次点源的夫琅禾费衍射，我们利用结像透镜将此夫琅禾费衍射成像在后焦面上，所以结像透镜的像是物镜傅里叶像的严格傅里叶变换。从另一方面来看，这种情况也可以看成是$d = \infty$时式 24 的近似，也是满足傅里叶变换关系。在这个过程中，厂家为了较好的成像质量，尽量将物镜出瞳径径做大同时将出瞳角限

制在较小的数值上，从而使得尽量多的信息传递到了结像透镜的入射面。

## 4 傅里叶后焦面成像前沿应用

傅里叶光学的理论在上世纪六十年代就已经非常完善，并且广泛应用了。与表面等离激元光子学的重新兴起及纳米光学的发展一样，傅里叶光学的现代重新发展得益于上世纪九十年代开始大规模普及的微纳加工技术。纳米光学的兴起使得在衍射极限的尺度有了新的值得研究的现象。由此我们一下介绍的傅里叶后焦面成像的前沿应用也主要聚焦在纳米光学方面。

### 4.1 空间滤波或 $k$ 空间重构成像

从本科课本上我们知道，传统傅里叶变换的一个直接应用就是进行空间滤波，并且直接进行图像微分处理，这也是阿贝尔原理的直接应用。相关的现代进展还是进行图像微分从而使得图像更加的清晰。

在光学课程中我们知道，当一个物处于物镜前焦面时，在傅里叶面上会出现很多衍射斑点。如果我们挡住高阶的衍射斑点，像面上的图像将变得很模糊。反之，如果我们挡住低阶的衍射斑点，那么像的边缘将变得非常锐利，就像在图像处理软件中进行的图像微分一样。这个原理在现在前沿研究中被用来进行可编程图像处理。如图 9 所示[7]，物体位于物镜前焦面，L1 是结像透镜，其所成的像在图中像面的位置。经过透镜 L2，可以在 L2 后焦面上成傅里叶变换像，可以在 LCOS 位置进行图像采集。在这项研究中，LCOS 前面是一个可编程空间调制器，可以通过编程实现不同位置，不同区域的光透过，或者被反射，即傅里叶面的空间滤波器。被反射的光经过 NPBS 的反射，再次通过透镜 L3 成实空间像。如图 9 所示相当于一个高通滤波器，可以进行图像微分，所成像也只有边界。

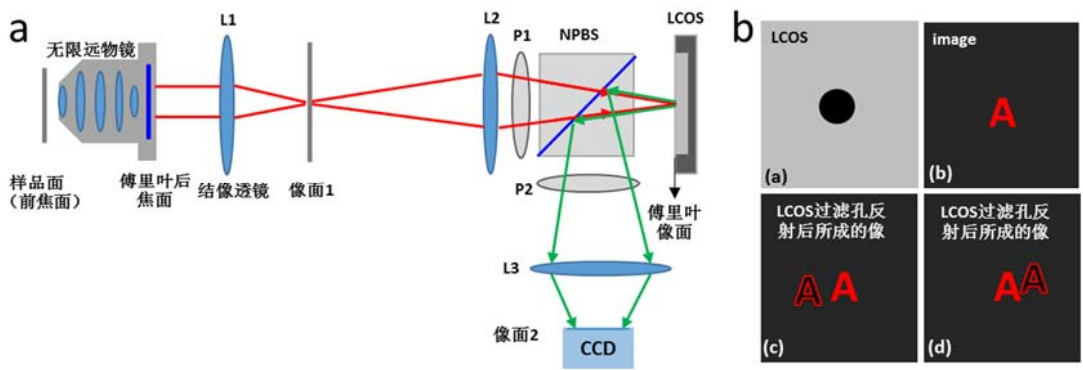

**图 9** （a）文献 7 所示的一种可编程傅里叶滤波成像显微镜示意图。（b）文献 7 所示的成像示意图：a.可编程空间反射调制器，b.样品的像；c,d 利用这种显微镜以及 a 所示的滤波孔进行调制以后在不同的偏振下所成的像[7]。

另外近年来发展出一种利用傅里叶面滤波进行图像重构从而实现高分辨成像的被称为傅里叶叶片瞳显微镜（Fourier ptychographic microscopy，FPM）的技术。这种技术一开始是利用计算机程序对图像进行傅里叶变换再重构[8]。但是最近 Tian 等利用物镜的后焦面位于物镜出瞳位置直接实现用出瞳进行傅里叶滤波，之后再利用算法算出出瞳透过函数及傅里叶空间

像从而进行图像重构[9]。如图 10 所示，LED 照明阵列可以发出光从不同的角度照亮样品。由光学位移相移定理我们知道这相当于在样品衍射单元上附加一个入射的相位差，这样夫琅禾费衍射斑点在 $k$ 空间会产生相应的位移。而出瞳孔径的位置不变，所以可以进行一定的滤波。不同的 LED 位置照明相当于在 $k$ 空间取不同的空间频率进行滤波，然后进行成像。在获得图像之后，利用傅里叶变换算法获得一定的振幅与相位信息，并且将此振幅与相位信息作为输入信息，再改变 LED 照明位置采集图像，进行下一轮的傅里叶变换，这种算法是自洽的，从而最终获得重构的图像。我们在此并不想将注意力聚焦到重构算法上，而主要关注光学傅里叶变换内容。在这种技术中之所以能够获得高分辨图像，用文献中的话说就是可以变相的提高数值孔径。如果利用我们第 2 节讲述过的知识，我们可以知道，当利用倾斜光照明时，相当于将 0 级衍射进行偏移，从而使得高阶的夫琅禾费衍射光进入透镜，同时高频信息及部分精细结构的信息也会进入透镜（图 10，倾斜照明示意图）。相似的通过计算机傅里叶变换重构的工作见 ref[8, 10-11]

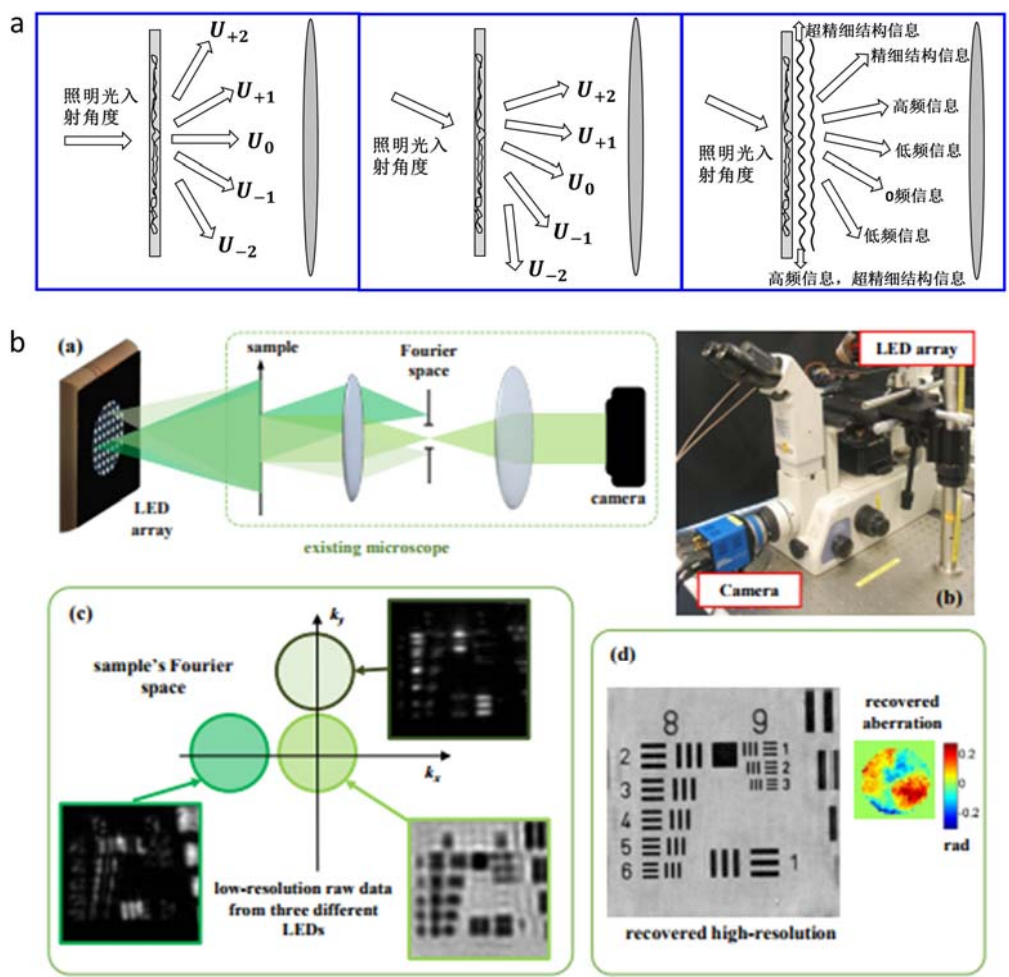

图 10 （a）不同倾斜角照明时的不同级衍射光方向示意图。（b）一种通过多点随机倾斜照明方式来获得高阶信息进行高分辨图像重构的显微镜；a. 原理图；b. 实验装置；c 不同位置 LED 照明时所对应的 $k$ 空间位置及所成物像；d. 最终所成的高分辨图像。(图片许可转载自参考文献 9，OSA)[9]。

**4.2 漏模显微镜及其应用：波导模式测量**

从第 2 小节的讨论中我们知道，傅里叶变换的物面上的空间频率 $f_x$ 实际上反映了衍射光在垂直于光轴方向传播的波矢（$k_x = 2\pi f_x$）。拿余弦光栅来理解，就是光波在焦面方向满足周期性边界条件，从而使得只有与此周期匹配的光波存在。从另一方面来看，只要焦面上存在周期性光波，并且能传播到远场，那么就能利用傅里叶显微镜进行变换成像。这种傅里叶后焦面成像装置（第三节所述的装置）在文献中一般被称为漏模显微镜（Leakage Radiation Microscopy, LRM）。这是由倒易原理来的。我们知道，当一束光从光密介质入射到光疏介质时，如果入射角大于临界角，将会发生全反射现象，并且在界面处产生隐失波，隐失波的波矢与入射光波矢的平行分量相同。反之，如果在界面上有一列传播的隐失波，则相应的，在光密介质中会在一定的角度产生一束出射光，出射光波矢的平行分量与隐失波的波矢相同。所以我们将这束出射光称为漏模（Leakage mode）。其原理如图 11a 所示。由第二小节的讨论我们知道隐失波的波矢实际上反映了界面上的空间特征，所以利用这个现象我们可以进行傅里叶变换成像。所以这种显微镜一般用倒置显微镜并且利用油浸入式物镜直接匹配样品的衬底的折射率。

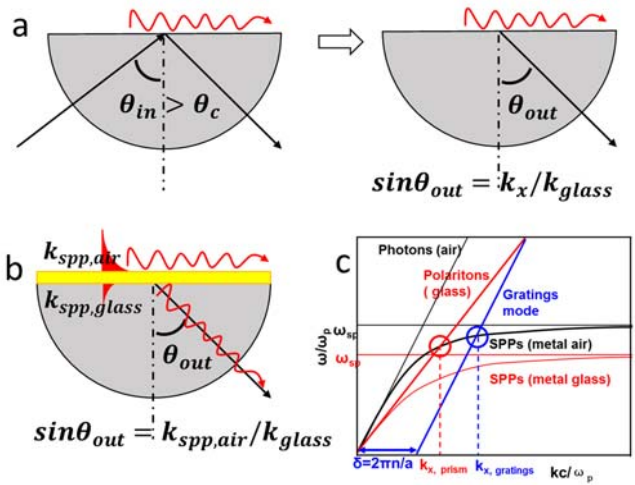

**图 11 光学倒易原理（a），金属薄膜 SPP 漏模（b）以及金属薄膜两个界面的色散关系 [12]（c）。**

目前对于波矢成像测量最普遍的就是对表面等离激元的波矢成像。其原理可以简单总结如图 11b 所示。位于玻璃衬底上的金或者银薄膜上可以产生传播的表面等离激元（波矢为 $k_{spp,medium} = \frac{\omega}{c}\sqrt{\frac{\varepsilon_{metal}\varepsilon_{medium}}{\varepsilon_{metal}+\varepsilon_{medium}}}$），我们暂且不论是表面等离激元的产生，就其传播来说，在空气/金属界面存在一列波（波矢为 $k_{spp,air}$），在金属/玻璃界面也存在一列波（波矢为 $k_{spp,glass}$）。其中 $k_{spp,air}$ 略大于空气中的光波矢 $k_0 = \frac{\omega}{c}$，$k_{spp,glass}$ 略大于玻璃中的光波矢 $k_{glass}$。根据我们之前所述的漏模的原理，$k_{spp,glass}$ 在这种平面结构中无法与玻璃中的波矢 $k_{glass}$ 的平行分量产生波矢匹配（图 11c 红线所示），所以是无法形成漏模的，只有空气/金属界面的波矢 $k_{spp,air}$ 可以与玻璃中的波矢（以一定的倾斜角）产生波矢匹配，可以产生漏模（图 11c 红圈所示）。当然只有膜很薄的情况下才可以产生漏模。当膜很薄的时候，由于上

下表面的波会产生相互作用，其共振能量发生一定的远离（$k_{spp,air}$能量增大，$k_{spp,glass}$能量减小），如果用固定频率的光激发，则相应的$k_{spp,air}$，$k_{spp,glass}$也会发生一点移动。当然，对于$k_{spp,glass}$来说，如果在金属/玻璃界面加工一定的周期性（周期为T）结构，使得其满足布洛赫条件，可以将其波矢等效的移动$\frac{2n\pi}{T}$（$n$为整数，图 11c 蓝圈所示），则也可以对其进行成像。当然根据具体实验方案，还有其他的配置，例如在金属薄膜上增加玻璃薄膜等。下面我们来看几个具体的研究例子。

早期的研究是直接利用半球透镜的发射方向在出射面不改变的原理来成像 [13-14]，所以不存在我们之前讨论的傅里叶变换关系，我们略过不谈。较早期的直接利用漏模显微镜的对波矢成像的工作是 Drezet 等人做出的 [15]。如图 12 所示，他们利用一个扫描近场光学显微镜或者一个单颗粒缺陷激发起表面等离激元，然后等离激元的漏模经过傅里叶变换，在后焦面除了直接入射的光斑之外，还存在两个半弧形的线条。那个就是漏模的波矢，其半弧的半径对于一定的漏模角度。半弧形表示表面等离激元波传播时成发散的球面波。

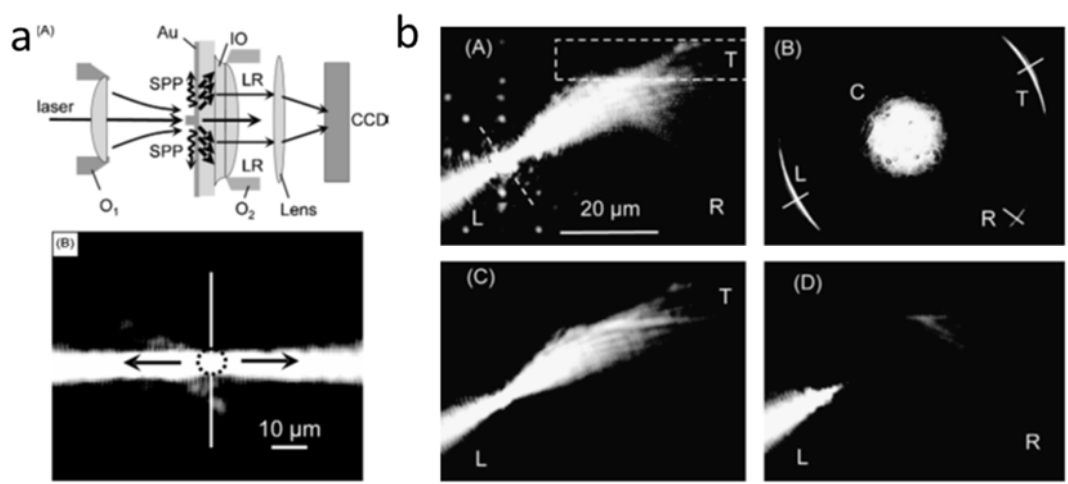

**图 12 利用 FBP 进行 SPP 传播方向及波矢的测量。（a）显微装置示意图及实空间传播方向示意图。（b）对应的实空间传播方向图(A,C)以及傅里叶像面图(B,D) (图片转载自参考文献 15)[15]。**

稍后比较典型的实验是 Berthelot 等人进行的 [16]。如图 13 所示，他们将一束激光打在一个薄膜的方形边角处，利用缺陷的散射可以同时激发起薄膜上的表面等离激元以及棱上的表面等离激元极化子。由于在膜上传播的 SPP 具有一定的发散角，类似于球面波，所以在傅里叶后焦面上成一个圆弧状。而激发起的边缘模由于只沿着y方向传播，所以在傅里叶后焦面上，是一条直线，只有y方向的值。在x方向相当于对一个点做傅里叶变换，所以在$k$空间应该具有所有的频率，所以是一条沿着x方向的直线，具有所有可能的$k_x$值。其他的工作还有[17-20]。

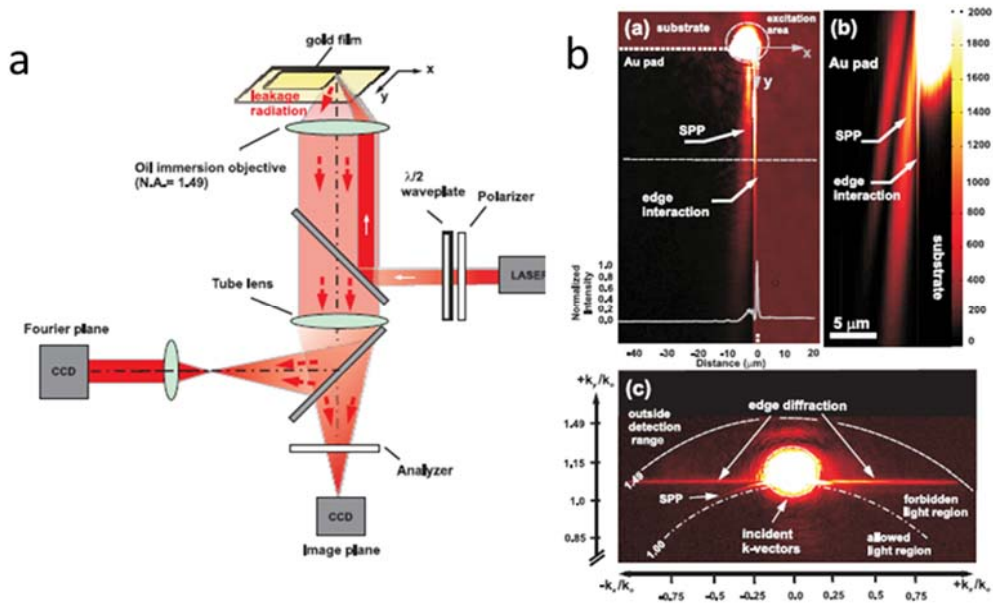

**图 13 利用 FBP 对不同 SPP 传播模式的波矢进行成像。（a）装置示意图。（b）实空间图像(a,b)以及对应的傅里叶面的图像(c) (图片许可转载自参考文献 16, OSA)[16]。**

其另外一个非常典型的应用是对一维纳米线波导进行波矢成像[21]。如图 14 所示，对于一个在金膜上的方形介质波导，对于不同的宽度，存在不同的波导模式（图 14a）。当利用波长为 800 纳米光激发不同宽度的波导时，会激发起不同的波导模式。利用傅里叶后焦面可以对不同模式成像（图 14b）。如图 14c 所示对于一个宽度为 1.5 微米的波导进行激发，则其存在$TM_{00}$模，$TM_{01}$模与$TM_{02}$模，可以在傅里叶后焦面上直接进行成像。一个更加清晰的例子请参见参考文献[22]。在这种周期性波导模式的成像中，高阶模归根结底是由于在波导横截面方向上存在驻波模式，所以在横向方向上也具有波矢，所以在成像中具有对称的波矢结构。周期性波导在不同的检测偏振下所成的 BFP 像，图像明显显示了不同的模式。另外，更加广泛的波导成像是用在表面等离激元金属纳米线波导波矢成像上。金属纳米线波导一个不同于介质波导的非常典型的特征是，对于基模不存在截止半径，而且基模的局域能力随着纳米线半径的减小而增大，所以可以突破衍射极限进行光信号传导，因此在现代光子芯片方面具有重要的应用价值。对于金属纳米线来说，其激发的模式与纳米线的直径及周围的介质折射率密切相关[23]。而且当把纳米线放在衬底上时，其模式会由于与衬底的相互作用而发生杂化，产生新的模式[24]。而且这种相互作用也使得我们可以利用漏模显微镜对其模式进行测量。如图 15 所示的银纳米线波导其直径只有 400 nm，当其进行导光时，我们在傅里叶后焦面能够明显看到其波矢像[25]。另外非常典型的应用见文献[26-27]。

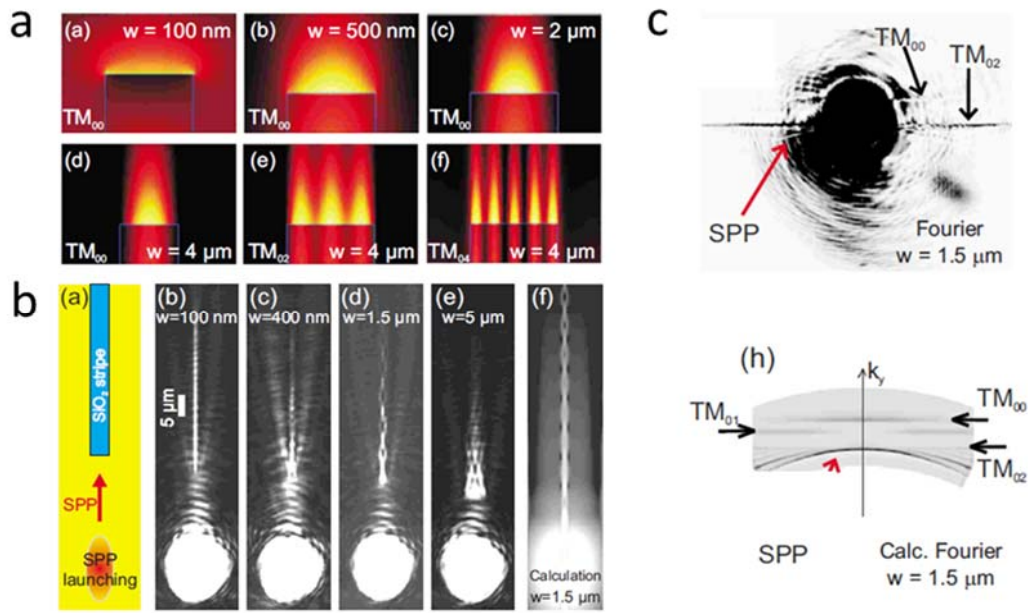

图 14 利用 FBP 对波导模式进行成像及测量。（a）置于金膜上的不同宽度的波导及对应的激发模式。（b）对不同的宽度的波导用 800 nm 光进行激发的实空间图像（a–f）及（c）宽度为 1.5 微米的波导的 FBP 所成模式图（g, h）(图片许可转载自参考文献 21，APS)[21]。

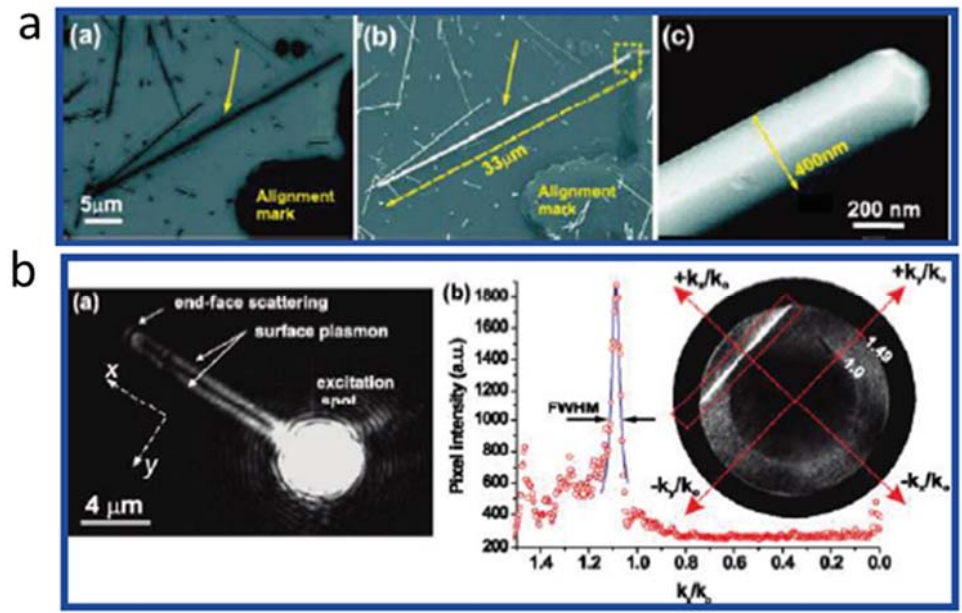

图 15 利用 FBP 对银纳米线波导进行测量。（a）纳米线明场图像，扫描电镜图像及局部放大的扫描电镜图。（b）波导实空间及傅里叶面成像(图片许可转载自参考文献 25，ACS)[25]。

### 4.3 漏模显微镜及其应用：色散关系测量

既然我们可以直接利用漏模显微镜测量波矢，那么我们就可以利用其进行色散关系的测

量。Giannattasio 等人其实在一开始直接利用 CCD 直接贴合衬底的方式进行色散关系测量。其利用的还是漏模原理，但是周期性结构的衍射光在经过有限的距离以后直接被 CCD 所接收（图 16），从第二小节的分析我们知道，这种配置其实不满足严格的傅里叶变换关系，还有一个不确定的相位因子。而且由于没有经过聚焦，在 CCD 上将会有较宽的条纹，不利于准确测量 [28]。

在 4.2 节我们看到利用傅里叶后焦面成像可以直接对波矢进行成像，那么如果我们同时利用不同的波长的光激发（或者单色光激发测量荧光），那么如果利用棱镜或者光栅进行分光，就可以直接对色散关系进行成像。如图 17 所示，Taminiau 等人利用一个狭缝限制，只取傅里叶面的一小块，然后将光路直接接入成像光谱仪，傅里叶像经过光谱仪光栅分光以后，可以直接在光谱仪 CCD 上成像，这个像由于狭缝的限制，在竖直方向上为波矢的值（横向波矢限制为 0），在横向上，是频率（或者波长）分布，由此直接成色散关系图 [29]。

Thomas 等人利用傅里叶后焦面成像对于不同周期的周期性一维光子晶体波导结构的色散关系进行了测量（图 18）[30]。由于周期性结构存在布洛赫模与不同的波长相匹配，所以当改变光子晶体的晶格常数时，会同时改变其波导内的共振模式及波矢，当用傅里叶后焦面成像时，其原理与之前所述的波导成像一样，不同的波导模式会直接成像在傅里叶后焦面上（图 18c），利用这些模式可以直接计算出波矢，利用晶格常数可以计算出响应的归一化频率。由此可以画出色散曲线来（图 18d）。

其他的无缺陷的二维周期性结构成像其实可以归结为对于光子晶体的波矢及态密度成像[31-32]，我们将在后面的小结讨论。

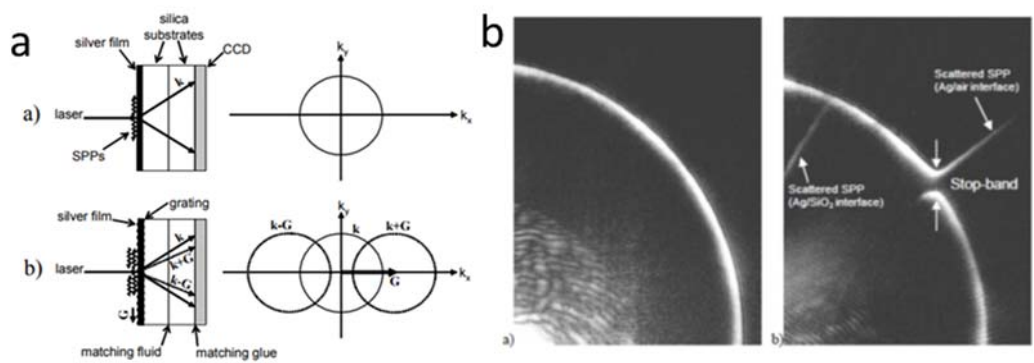

**图 16 利用 CCD 直接对漏模进行色散成像示意图（a）及所成像图（b）(图片许可转载自参考文献 28，OSA)[28]。**

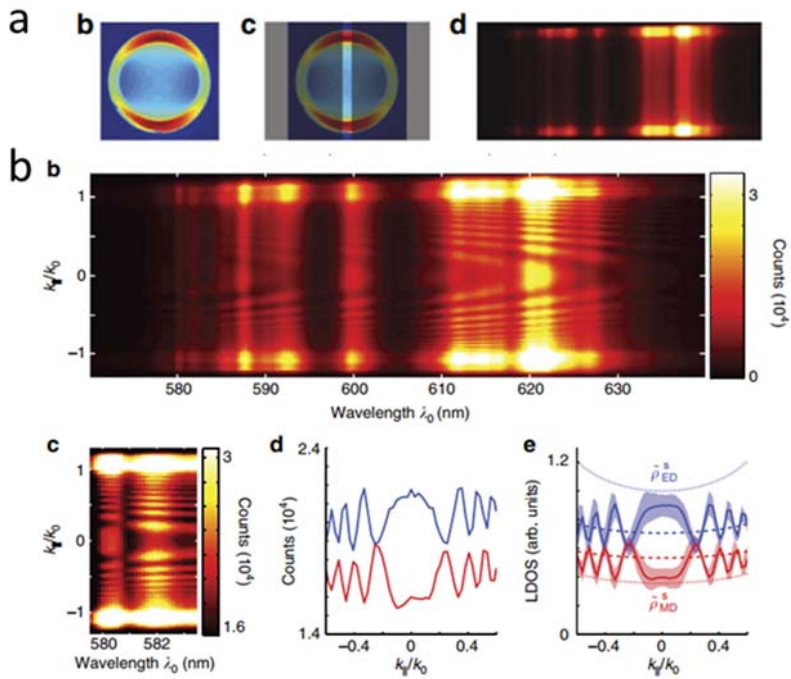

图 17 一种色散关系的测量方法(图片许可转载自参考文献 29，Springer Nature)[29]。（a）在傅里叶面上利用一个狭缝挡住其他方向，只留竖直方向的像，然后利用成像光谱仪所成色散图。（b）一种样品的色散关系图。

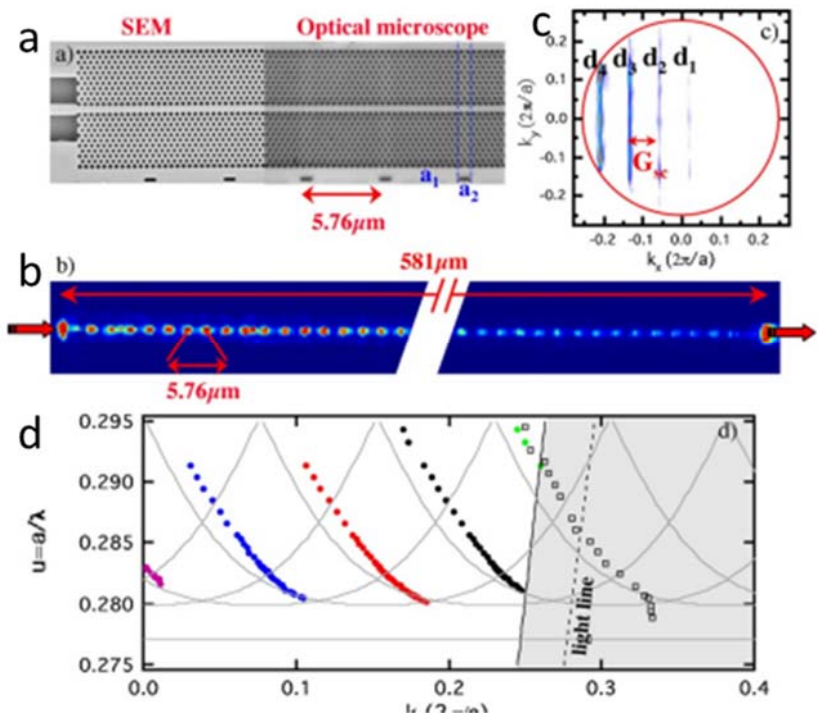

图 18 利用 FBP 测量色散关系。（a）基于 GaInAsP/InP 的光子晶体波导。（b，c）利用1535 纳米光激发波导的实空间及 FBP 像。（d）第一、第二、第三及第四布里渊区色散曲线图(图片许可转载自参考文献 30，OSA)[30]。

**4.4 发射方向测量**

傅里叶漏模显微镜的重新流行是 2004 年 Lieb 的工作所引发的。在他们的工作中，主要用傅里叶后焦面成像来进行单分子发射方向的测量[1]。如图 19 所示，从直观上来讲，只要将点光源置于透镜焦面上，在另外一面就可以直接测量到发射方向的分布。稍早一些的工作是直接将点发光源置于半球状透镜的平面上，由于光在半球边界折射角为 0，可以直接反映光源的发射方向。但是我们需要从傅里叶变换的角度来看这类研究，而且，现在的点光源如分子或纳米颗粒偶极子及多极子发射方向的研究都用傅里叶显微镜来进行。

最早这方面的工作是在二十世纪初由 Sommerfeld 等人做出的平行或者垂直于界面的偶极子发射分布[33-34]。之后 W. Lukosz 与 R. E. Kunz 在 1977 年至 1979 年发表的三篇非常经典的文献中详细计算了位于距离界面非常小的电偶极子以及磁偶极子发射强度及发射空间角分布[35-37]。在他们的计算中，尽管偶极子距离界面非常近（$z \leq \lambda_1$），但是他们还是利用偶极子自由空间的发射分布以及菲涅尔系数来进行计算，透射分布即为按照菲涅尔系数进行计算的，反射分布为远场条件下偶极子的直接发射与界面反射的相干叠加。事实证明这种计算结果还是非常准确的。稍后 Lieb 等人实际利用了同样的做法，但是没有考虑更深层的物理原理，直接得出了与 W. Lukosz 一样的结论，不过进一步利用傅里叶后焦面成像证明了这种空间分布结果[1]。这种单个分子发射强度的空间角分布，其实也可以直接从傅里叶变换光学的角度来看。事实上在 W. Lukosz 的理论中，由于光从偶极子的位置照射整个界面平面时，由于入射角的不同，其波矢的平行分量也不同，所以在进行这些计算时他首先就对偶极子的标量势在界面进行了傅里叶分解，然后对不同波矢的光独立进行菲涅尔系数的计算[37]。我们在此也可以利用上一小节隐失波的概念来理解。我们知道事实上偶极子是处于界面半波长以内的，所以偶极子的发射完全可以激发起界面处的隐失波。隐失波的波矢由边界条件所确定，取决于偶极子的偏振方向和与边界的距离。而根据倒易原理，这个隐失波可以产生相应的漏模，从而傅里叶变换事实上是对隐失波空间频率的一个变换。这样，利用傅里叶显微镜对点光源发射方向成像就与之前的理论统一了。**从另外的角度来讲**，处于透镜焦面的一个点的傅里叶变换应该是对应整个实空间的平面，而不是某些实空间的点。点偶极子的发射空间角分布在物镜的入瞳面来看，是已经经过了菲涅尔衍射分光以后的结果，这些分光光场经过物镜以后，就其光场来看，在其后焦面也是可以近似理解为发射源的傅里叶变换的。

在点光源发射方向方面的工作算是利用傅里叶后焦面成像最普遍的了。在近年来引起热潮的 Lieb 的工作中（图 19）[1]，他利用倒置显微镜搭建的傅里叶后焦面成像装置对荧光分子进行了单分子实空间及傅里叶后焦面成像，利用图 19b 所示的对应关系可以直接利用傅里叶后焦面的强度分布计算出分子在实空间的取向。之前 4.2 小结提到的 Hartmann 等人的工作尽管是利用傅里叶显微镜对表面等离激元波进行成像，但是由于其利用单根碳纳米管激发 SPP 波，所以也能根据波矢方向来确定碳纳米管的取向[27]。另外一个非常典型的工作是 Curto 等人做出的[38]。如图 20 所示，一个金属纳米棒的局域表面等离激元共振由于存在偶极与高阶的激发模式，其发射空间角分布会产生像电偶、电四、电八、电十六极矩等能量分布。同上面在衬底上的单分子偶极子的发射方向一样，利用傅里叶显微镜可以对金属纳米棒的多极子（此工作中利用量子点进行激发）的发射方向进行成像（图 20c-n）。

如果我们考虑两个平行方向放置的相距一段距离的偶极子，则这个体系的远场发射方向

会取决于两个偶极子辐射场的相干叠加[39]。如图 21 所示，两个置于玻璃衬底上的金和银的纳米圆盘（直径分别为 130 纳米和 110 纳米）相距 15 纳米。由于金和银纳米盘的局域表面等离激元共振波长不一样，金的大约在 660 纳米左右，银的大约在 550 纳米左右，其激发如同弹簧振子的振动模式激发一样，如果激发波长小于共振波长，则圆盘的响应比激发落后 π 相位，如果激发波长大于共振波长，则圆盘的响应与激发同相位。所以利用不同波长的光同时激发这两个圆盘，则其发射方向也会有不同的分布。而如果用白光激发，则绿光与红光由于相干叠加会被散射到不同的方向。利用这个原理，可以直接对颜色进行分光。而利用傅里叶后焦面可以直接测量这种分光现象。同时由于金属纳米盘的共振波长与周围环境的介电常数相关，所以当周围环境变化时，其共振波长的变化将会导致纳米傅里叶面上分光强度的变化。利用这种特性可以做成传感器，通过直接测量傅里叶面上左右两边光强的比值，从而判断是否有响应的分子吸附（图 22）[40-42]。

除此之外，还可以利用傅里叶后焦面对纳米线波导终端的发射方向直接进行成像，从而对于不同直径的金属纳米线波导，由于有效环境折射率的不同以及直径的不同，其不同的波矢可以从傅里叶后焦面直接测量出[43]。 其他应用见[44-45]

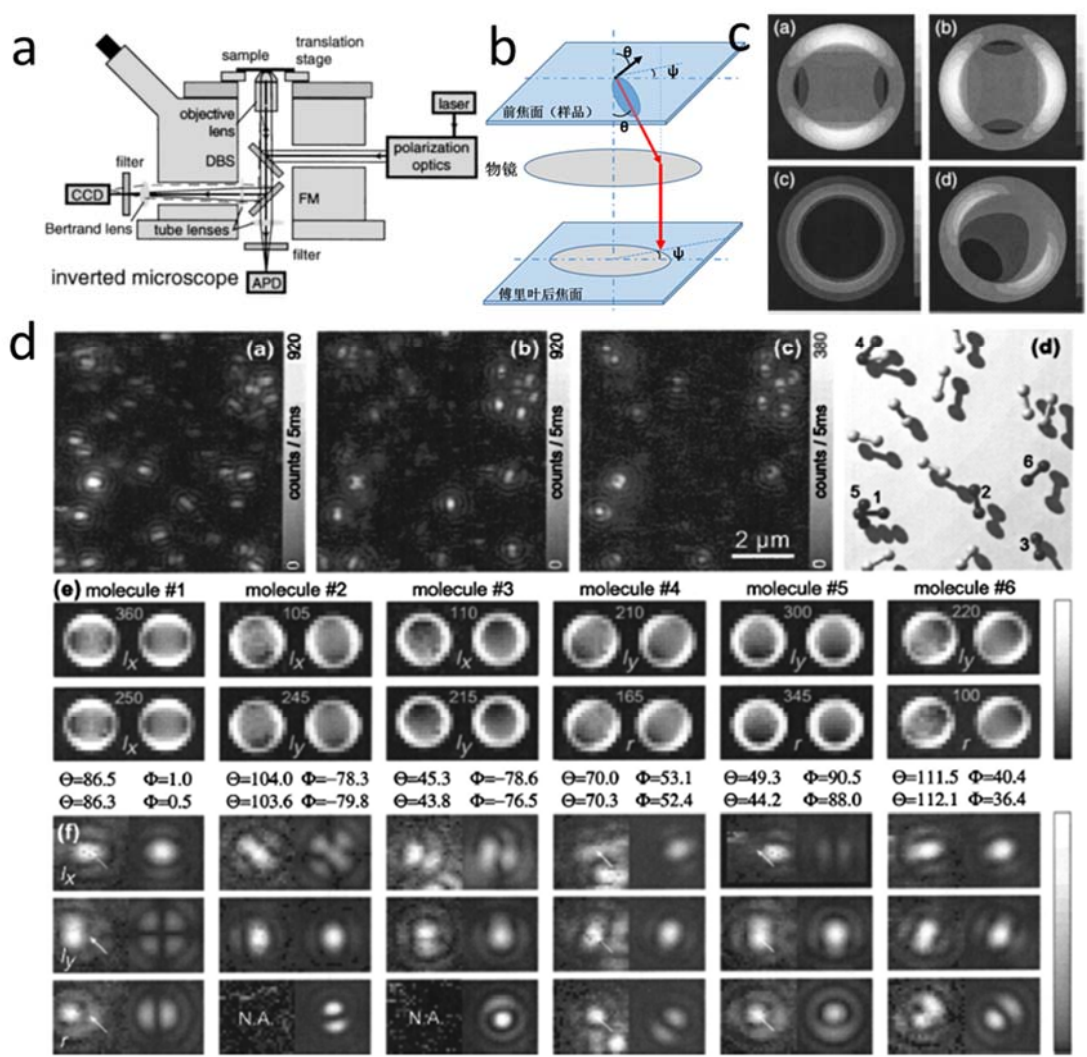

图 19 利用 FBP 对单个分子的取向进行成像。（a）装置示意图。（b）发射方向成像示意图。（c）单个分子不同取向的傅里叶后焦面图像。（d）不同分子的傅里叶后焦面图像以及计算出的响应的取向(a, c, d 图片许可转载自参考文献 1，OSA)[1]。

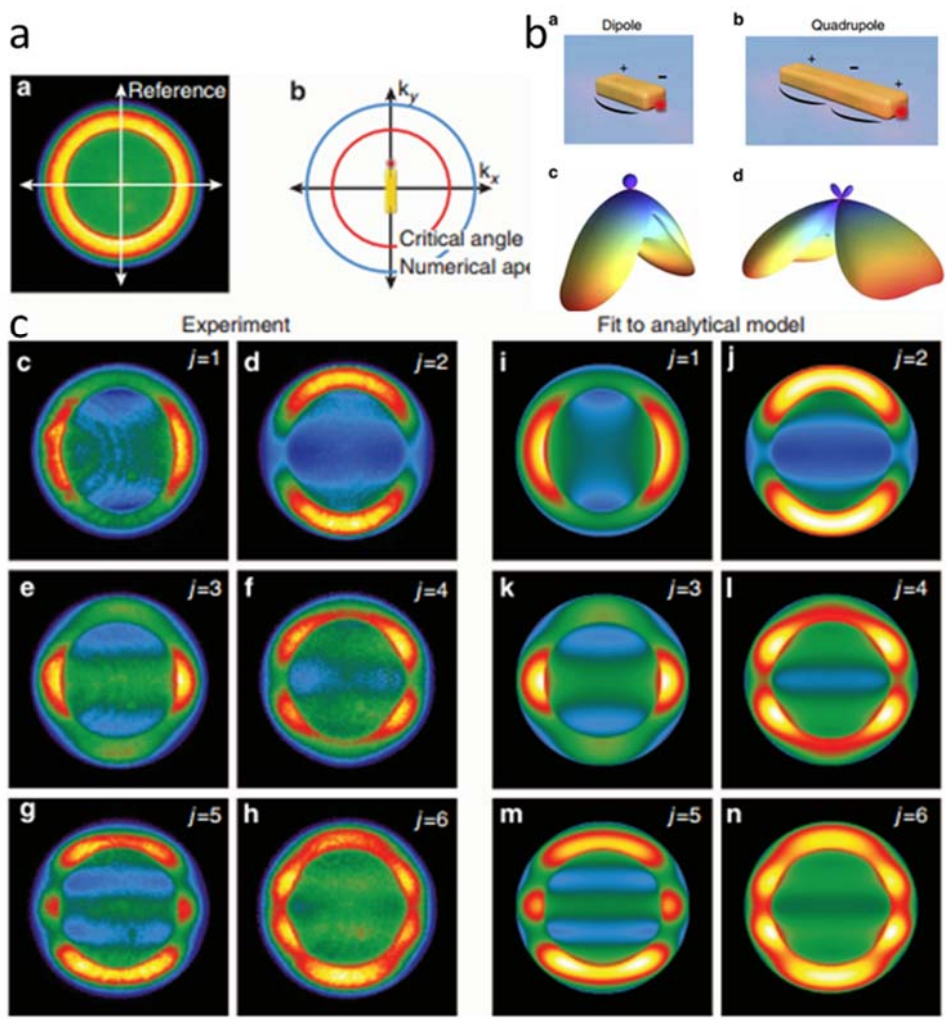

图 20 利用 FBP 对金属纳米棒的多极辐射进行成像。（a）金属纳米棒激发示意图。（b）置于玻璃衬底上的金属纳米棒偶极子及四极子激发及发射角度模拟图。（c）不同阶激发的纳米棒的 FBP 实验测量图及模拟图(图片许可转载自参考文献 38，Springer Nature)[38]。

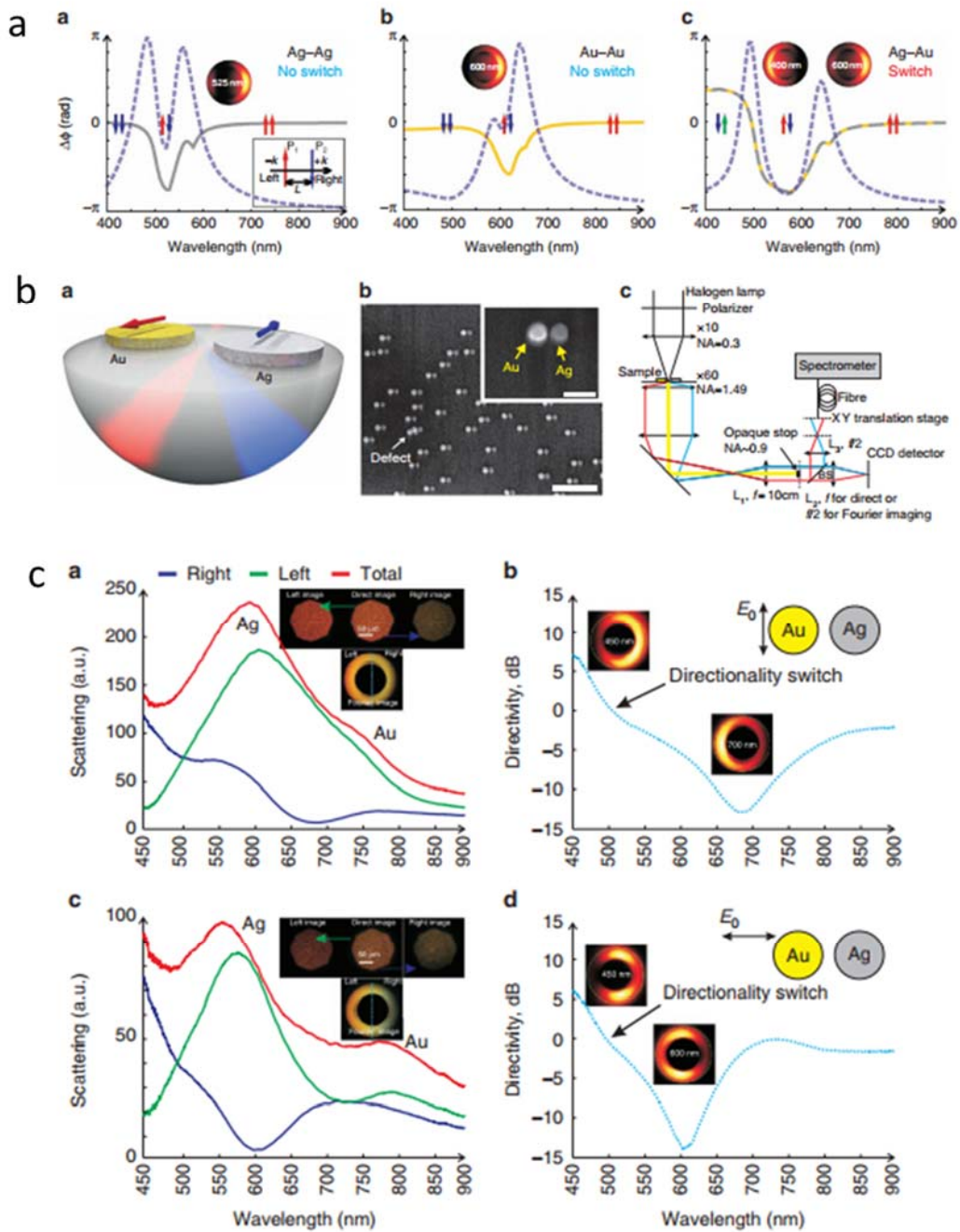

图 21 利用 FBP 对两个不同材料组成的纳米盘体系进行发射分光测量。（a）相同材料的两个圆盘，以及不同材料的两个圆盘共振是两个圆盘上振荡相位曲线。（b）金-银纳米圆盘结构在不同波长的光激发时的发射方向示意图、实验样品以及测量装置示意图。（c）上图中的样品在不同偏振下利用白光激发时 FBP 左右两边测量的光谱以及发射方向对比图(图片许可转载自参考文献 **39**，Springer Nature)。**39**

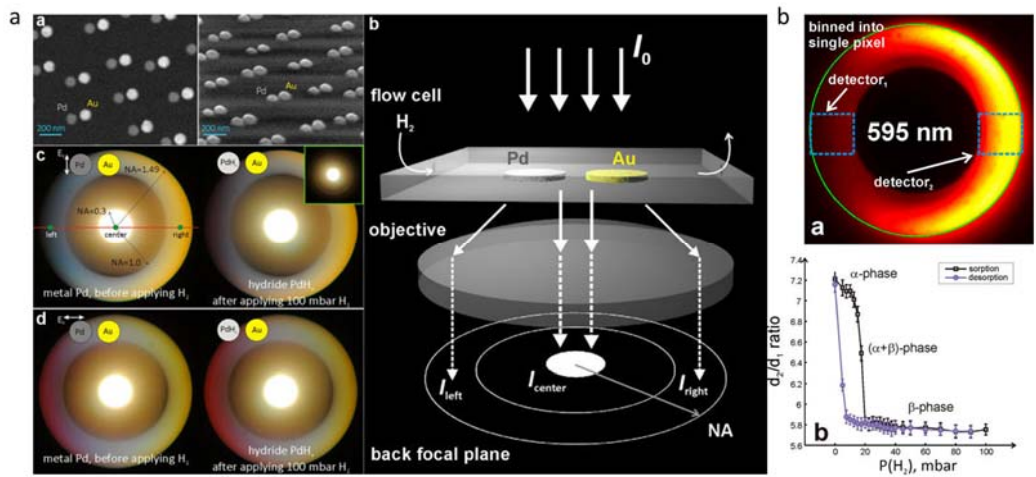

**图 22 利用双材料金属盘分光效应进行的传感应用。（a）样品图，FBP 示意图以及 FBP 实验图。（b）在 595nm 测量是对于氢气吸附传感实验(图片许可转载自参考文献 40，ACS)。[40]**

### 4.5 $k$空间成像及二维光子晶体

对于光子晶体的$k$空间态密度的成像，看起来和前面介质波导的成像很类似，就是对不同的波矢进行成像，但是这个波矢的形成是周期性结构导致的周期性的布洛赫波函数形成的，而不是之前所述的一个平面上传播的波的波矢。傅里叶变换光学，从直观上来讲我们也知道其可以将实空间信息变换为$k$空间信息。前面的推导用的是空间频率，但是我们知道由于布洛赫条件的，空间频率正好匹配$k$空间波矢。所以我们利用傅里叶显微镜可以直接对光子晶体等进行$k$空间成像。由此，对于$k$空间的成像其实更近于傅里叶变换的直观图像 [46-47]。

前面所述的图 14 的色散关系的测量其实也算一个例子 [28]。现在我们首先来看其对只在一个方向有周期性分布的样品的成像（图 23a）[48]。其在$x$方向是无限延伸的，在y方向上具有周期性的结构。制作是首先在衬底上面蒸镀一层金膜，然后在上面加工周期性的 PMMA 矩形条波导。其中 PMMA 中掺了 R6G 分子，用于荧光成像。然后利用激光激发样品，则在 PMMA 中会形成波导模式，同时在 PMMA 与金膜界面处也会存在 SPP 模式，利用 FBP 可以直接对波导模式进行成像，我们之前已经讨论过。另外一个非常显著的现象是在$k_y$方向上也存在波模式的平移，这个其实是布洛赫波造成的（$k_{diff,\parallel} = k_{inc,\parallel} + G$，$G = \frac{2\pi}{a}n$，$n$ 为整数）。更加显著的现象可以从如图 23a(b)所示的样品中测得。其金膜上面首先沉积了铝的周期性结构，然后再沉积 PMMA。这样形成了更加突变的周期性便捷条件，从而使得布洛赫波矢的移动更加明显。

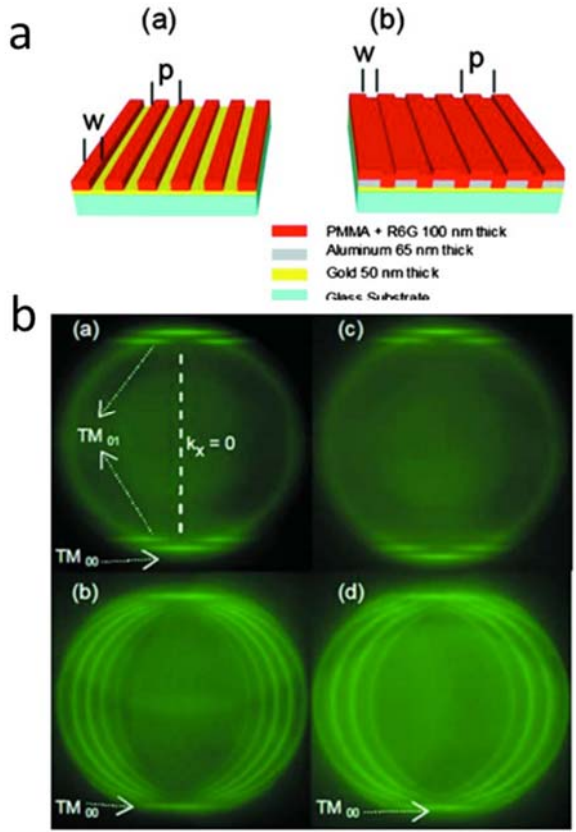

**图 23 利用 FBP 对一维周期性结构进行成像。（a）样品示意图。（b）对（a）中的 b 样品在不同偏振下的测量结果（竖直偏振为第一行，水平偏振为第二行）(图片许可转载自参考文献 48，AIP)**[48]。

作为傅里叶变换装置，其一大优势就是可以直接对光子晶体进行$k$空间态密度成像。如图 24 所示的样品[49]，玻璃衬底上沉积有 50 纳米的金膜，然后在金膜上加工二维周期性 PMMA 结构，PMMA 中掺有 R6G 分子用于荧光成像。样品利用激光直接激发，然后对荧光发射波长进行成像。荧光分子在周期性结构上形成 SPP 布洛赫态，然后通过漏模进入物镜。在傅里叶后焦面上我们可以直接看到对应的光子晶体的费米面。如图 24b 所示，内圈淡颜色的圈表示一个平均的有效介电常数对应的表面等离激元波，其波矢由样品周期决定。外圈表示物镜的数值孔径大小，及所成像的最大波矢值。在$k$空间也可以同时画出第一布里渊区（图 24b 白框所示）。由于周期性结构导致的平移对称性，在$k$空间波矢可以平移$G = \frac{2\pi}{a}n$。决定于$a$的大小，其有可能与第一布里渊区重合，其波矢方向与原波矢相反，所以有抵消作用，这时是个很暗的条纹（小圈中的暗弧线）。从图中我们看出其图案形状与固体物理中的费米面非常相似。当放大时，也能够看出小的带隙结构（同时见图 16b）。当改变晶格形状时，会改变布里渊区的形状时，从而改变费米面的分布（图 24b 右侧）。当改变晶格周期时，布里渊区的大小也随之改变，从而使得费米面的分布发生变化（图 24c）。我们知道在固体物理中态密度可以从色散关系计算出来，当色散曲线变的比较平时，对应某个频率的态密度会增加。我们之前也知道利用 FBP 可以进行色散关系的测量。对于晶体来说，比较标准的导出态密度的方式是在倒格空间中利用周期性边界条件来计算。但是利用 FBP 成像时，只要我们针对某一

频率（能量）的光直接进行成像，那么就可以直接得到 DOS 在 $k$ 空间的分布。直接进行积分就能得到 DOS（对于周期为 $L$ 的二维结构来说，波矢可以表示为 $\boldsymbol{k} = n_x \frac{2\pi}{a} \hat{\boldsymbol{x}} + n_y \frac{2\pi}{a} \hat{\boldsymbol{y}}$。在 $\boldsymbol{k_x}$ 与 $\boldsymbol{k_x}$ 面上，一个模具有 $\frac{4\pi^2}{a^2}$ 的二维体积）。更多相似的分析见 ref[50]

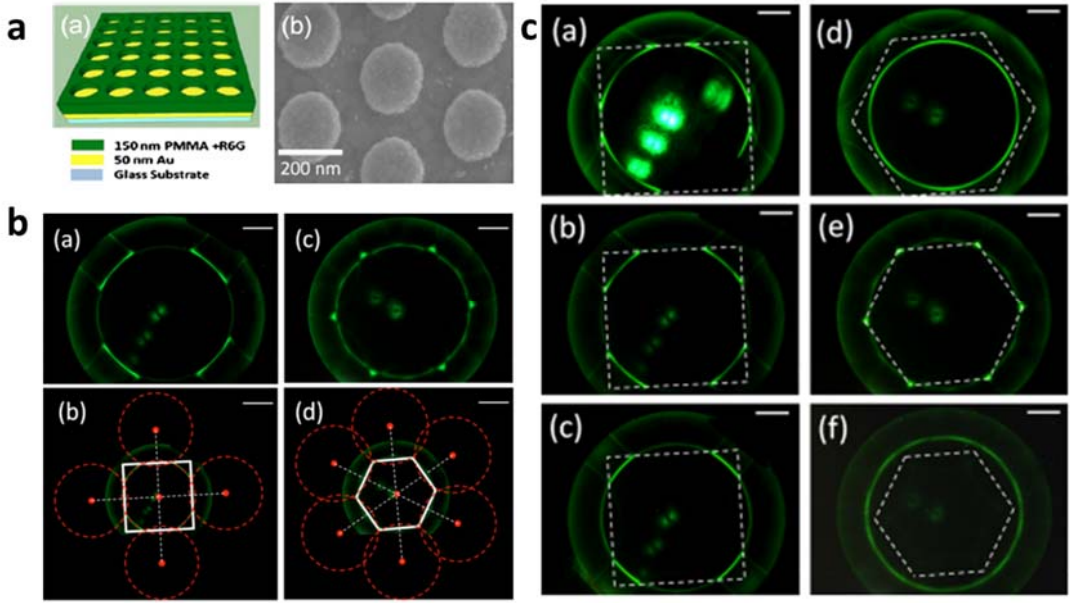

图 24 利用 FBP 对二维光子晶体费米面进行成像。（a）样品示意图，一种样品扫描电镜图。（b）对上面两个不同结构的样品的 FBP 所成的像（a,b 对应左边样品，c,d 对应右边样品）。（c）对上面两个不同结构的样品改变周期时所成的 FBP 像(图片许可转载自参考文献 49，AIP)[49]。

　　利用傅里叶后焦面成像不但可以对周期性光子晶体结构进行研究，也可以对准晶结构的光学性质研究。我们知道晶体结构的排列具有平移对称性及旋转对称性，非晶结构具有长程有序的结构但是不具备平移对称性。晶体结构具有 1、2、3、4、6 五种旋转对称轴。准晶结构是介于晶体与非晶之间的一种结构，其具有 5 重或者大于 6 重的旋转对称轴。由于其独特的性质，对准晶结构的研究工作获得了 2011 年诺贝尔化学奖。如图所示是由银纳米盘在玻璃衬底上组成的 Penrose 准晶图案。这种图案是由两个菱形的结构拼接而成的，其没有平移对称性，但是具有长程有序结构，具有 5 次对称轴。其傅里叶变换斑点显示出 10 重对称性。这种准晶等离激元结构一般没有近场或者远场的相互作用，其消光性质与单个颗粒基本一样。但是其傅里叶后焦面图像却表现出非常明显的有序结构。如图 25 所示，利用不同的光对其进行成像时，傅里叶图像表现出明显的 10 重对称性。随着波长的增加，衍射图案不断的远离中心斑点[51]。图 25b 中心的团是不同色光的衍射叠加。我们知道面内平行于界面的光波矢满足 $\boldsymbol{k_{diff,\parallel}} = \boldsymbol{k_{inc,\parallel}} + \boldsymbol{G}$，$\boldsymbol{G} = \frac{2\pi}{a}\boldsymbol{k}$，$\boldsymbol{k}$ 由五重对称性 $e^{i\pi/5}$ 的线性组合给出。$\boldsymbol{a}$ 是第一赝布里渊区的周期。在同一波长成像时，如果选择不同的曝光时间，则可以清晰的看到不同级 的 衍 射 图 案。 这 些 图 案 利 用 光 学 位 移 相 移 定 理 可 以 直 接 计 算 出 来（$I(\theta,\phi) =$ $F(\theta,\phi)S(\theta,\phi) = F(\theta,\phi)\sum_{i,j} e^{iq\cdot r_i}e^{-iq\cdot r_j}$）。我们看到计算的衍射图案与傅里叶后焦面成像的图案吻合非常好。其他的利用 FBP 对不同准晶图案的研究可以参见[52-53]。

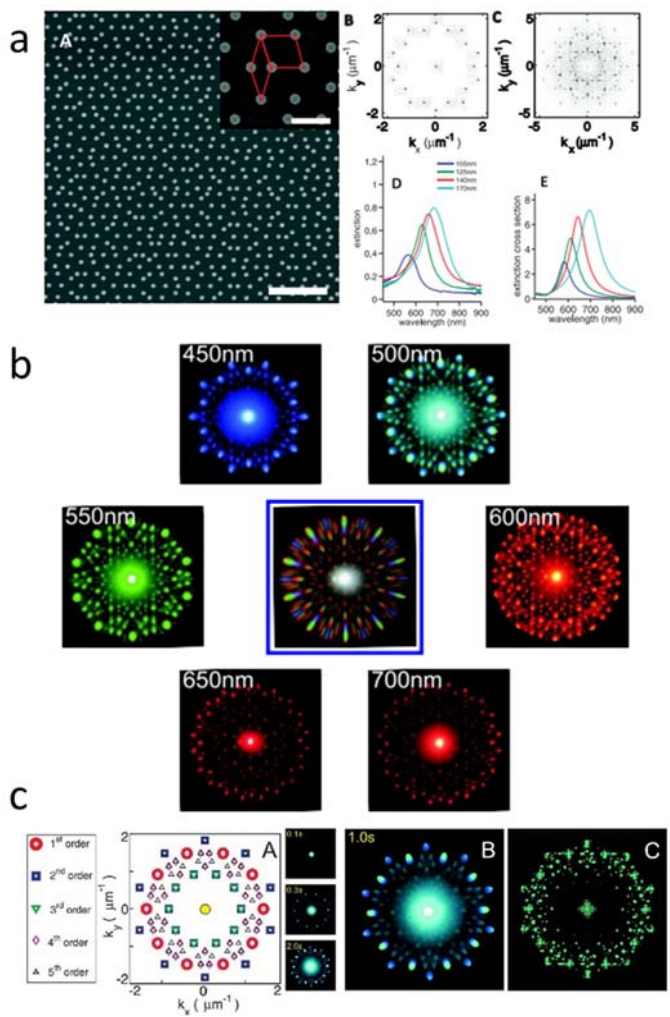

图 25 利用 FBP 对准晶的布里渊区进行成像。（a）微加工制作的金属纳米盘在玻璃衬底上的准晶分布图案及其数值傅里叶变换图，以及不同直径的纳米盘的消光谱。（b）利用不同的滤波片所成的 FBP 图像，中间的是白光成像。（c）在不同曝光时间下不同衍射级的 FBP 图像(图片许可转载自参考文献 51，ACS)[51]。

## 5 在本科实验教学中的相关实现

考虑到无限远光学显微镜都比较贵，一般的教学实验室又需要很多套设备，为了降低成本，我们可以在传统的低成本显微镜上来实现。

如 26 图所示，我们可以设计一个一体化的小管，内置波特兰透镜，一端可以直接插入目镜接口，这种接口可以直接通过网络廉价购买，然后接具有波片切换装置的管，在切换拉杆的两个孔上一个放入焦距为 $f$ 的波特兰透镜，另一个装入焦距为 $f/2$ 的波特兰透镜。然后在另一端可以直接安装一个照相机 CCD（网购很便宜），调整切换拉杆其处于显微镜像面与 CCD 的中间位置并且保证显微镜像面与 CCD 芯片距离为 $2f$，通过切换波特兰透镜，就可以同时进行实空间或者 $k$ 空间成像。由于现代科研显微镜的结构透镜及管长与传统显微镜兼容，所以这种装置可以直接插在传统显微镜上，或者科研无限远显微镜上进行成像。样品可以用

光盘等简单的周期性结构。这样物理教学实验室也可以用低成本的办法实现前沿科技的研究装置，更加理解前沿研究。类似的装置也可以设计为大物实验课的一次实验内容。

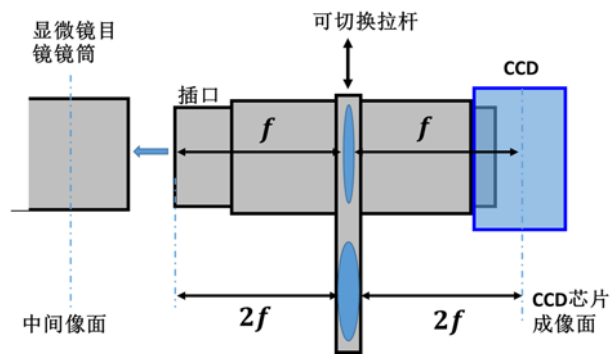

**图26 教学实验室简单傅里叶实空间成像装置设计。**

## 6 总结及展望

本文从菲涅尔衍射与夫琅禾费衍射的傅里叶分光本质以及薄透镜的相位变换出发，结合近代无限远光学显微镜所满足的傅里叶变换的条件，介绍了傅里叶后焦面成像装置及其在前沿研究中的具体应用。通过这些研究的介绍我们可以感受到课堂知识在前沿中是如何体现的。在最后我们发现这些前沿的技术也可以在物理实验课上相对容易地实现。

基于传统光学的傅里叶变换光学在上世纪六十年代成熟以后，在近年来由于微纳加工技术的飞速发展，又开始在纳米光学领域重新焕发出勃勃生机。由于纳米光子学以及量子光学的发展，我们相信在傅里叶光学成像技术将继续在我们文中所述的领域发挥巨大的作用。而且更进一步，我们有理由相信，在不久的将来，傅里叶光学这种基于简单透镜变换即可实现复杂变换运算的能力也将在量子运算及量子信息处理、并行光子芯片等方面发挥巨大的作用。而且如果我们能够在微纳尺度加工出具有同样功能的纳米透镜、实现微纳尺度的傅里叶后焦面成像，这项技术也许将实现一个质的飞越。

本文是一篇基于本科光学知识点的前沿综述，所以在前沿工作的介绍中难免会有一些浅显，不过并不阻碍对于傅里叶光学部分知识的理解。